# A closed-form solution for the flat-state geometry of cylindrical surface intersections bounded on all sides by orthogonal planes

Michael P. May

Dec 12, 2013

Presented herein is a closed-form mathematical solution for the construction of an orthogonal cross section of intersecting cylinder surface geometry created from a single planar section. The geometry in its 3-dimensional state consists of the intersection of two cylindrical surfaces of equal radii which is bounded on all four sides by planes orthogonal to the primary cylinder axis. A multivariate feature of the geometry includes two directions of rotation of a secondary cylinder with respect to a primary cylinder about their intersecting axes. See Figs. 1 and 2 for the definition of the two variable directions of rotation of the secondary cylinder with respect to the primary cylinder.

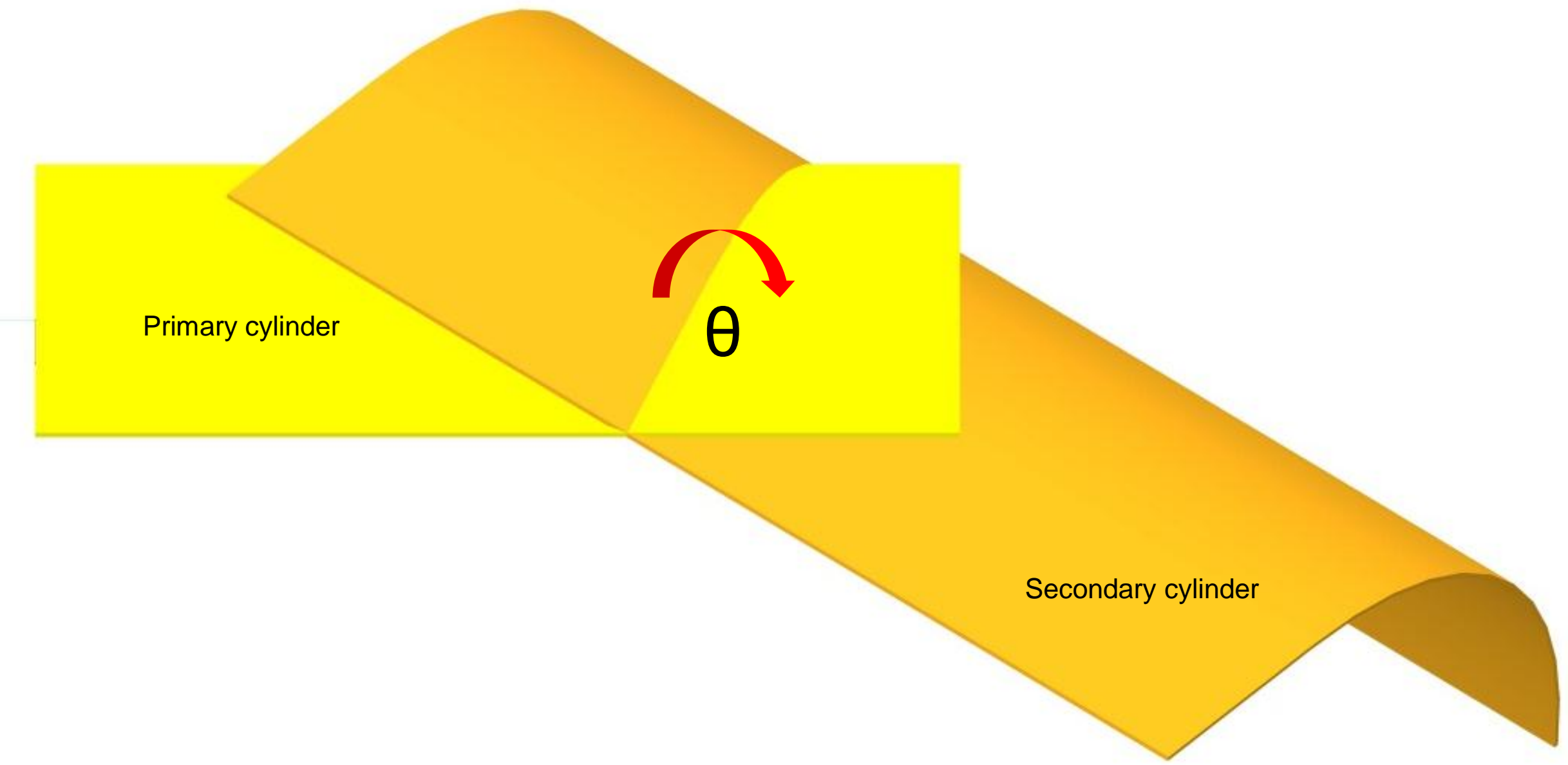

Fig 1: Primary direction θ of the rotation of the secondary cylinder section with respect to the primary cylinder section (top view)

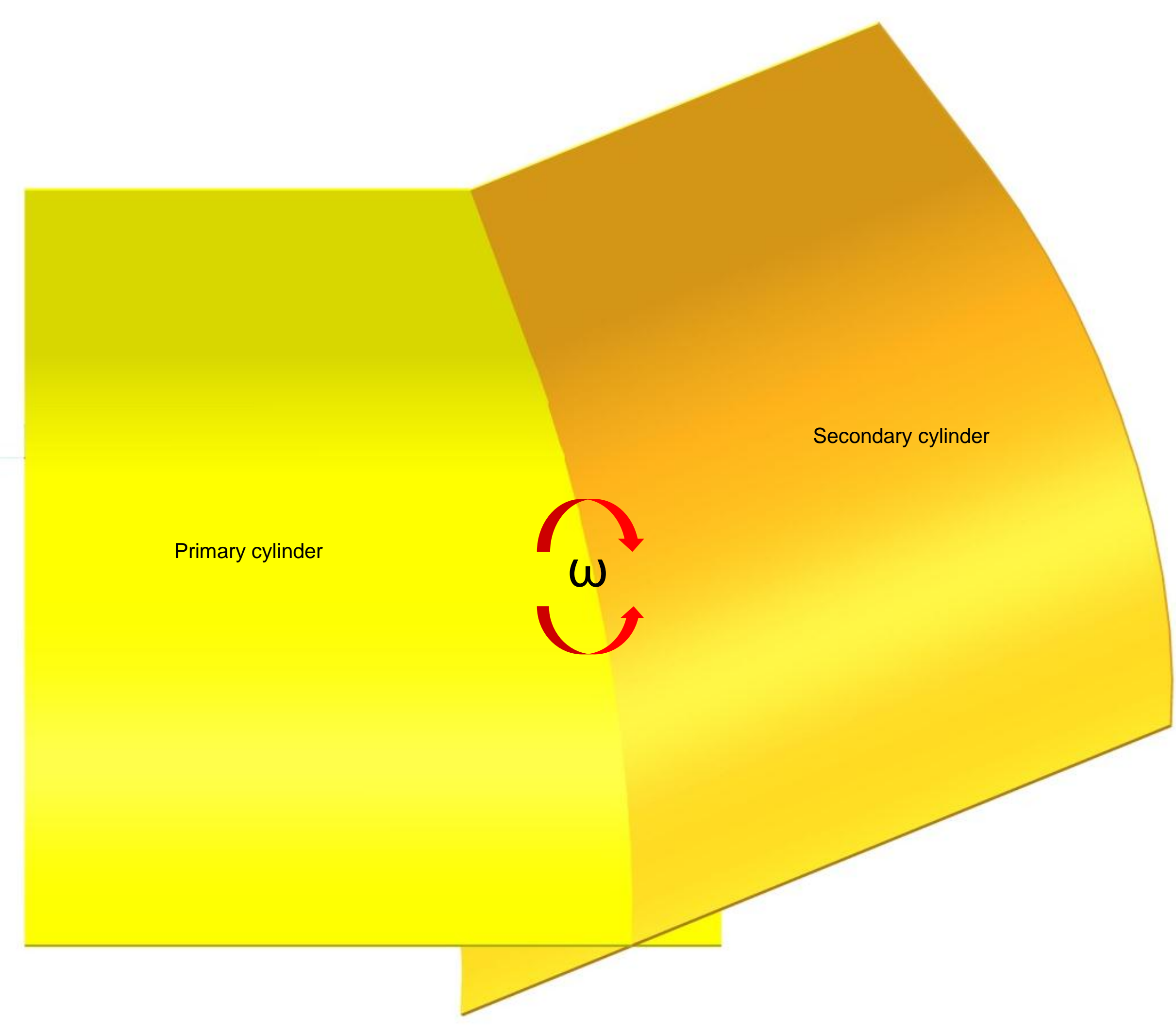

Fig 2: Secondary direction ω of the rotation of the secondary cylinder section with respect to the primary cylinder section (front view)

For convenience in forming the 3-dimensional geometry from its plane state, each cylindrical surface section will include a short section of planar wall on one side of the cylinder surface that is tangent to the cylinder along its longitudinal axis. It will later be shown that this short section of flat wall is expedient for producing the three-dimensional geometry from its flat state via rolling and bending operations that follow the profiling of the geometry in its flat state. See Fig. 3 for a visual depiction of the short planar wall sections that are tangent to each cylinder along one side.

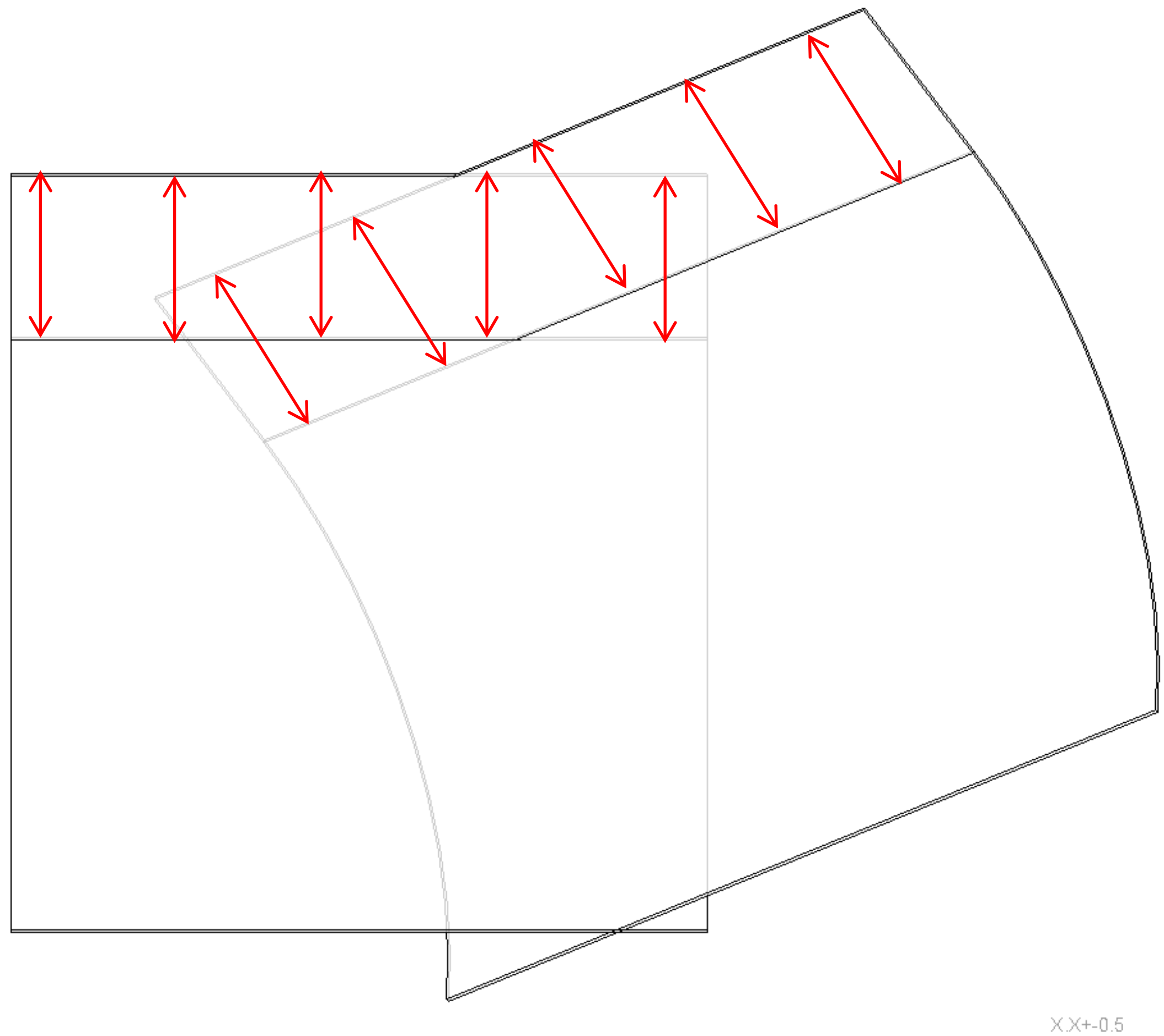

Fig 3: Planar wall sections tangent to the cylindrical surface sections

We will now proceed with an introduction of the cylinder surface intersection geometry. A distinguishing feature of this multivariate geometry is that it is formed from a single flat sheet. All the cuts, rolls, and bends necessary to produce the geometry are derived from sinusoidal curve equations with parameter inputs provided by trigonometric calculations created in a spreadsheet. The sinusoidal curve equations that are used to produce the cuts in the one-piece sheet to form the 3-dimensional geometry were borrowed (with permission) from the work of Tom Apostol and Mamikon Mnatsakanian in their document titled, Unwrapping Curves From Cylinders and Cones, which is published and available online. All calculations necessary for producing the parameters required of the curve equations and for creating the rolls and bends that form the flat state of the geometry into the 3-D state were derived by the author for application in this geometric design.

As an introduction to the challenges presented in constructing the variable geometry defined herein from a single planar section, please refer to Fig. 4 for views of the planar profile before, and the fully-formed geometry after, the rolling and bending operations are completed.

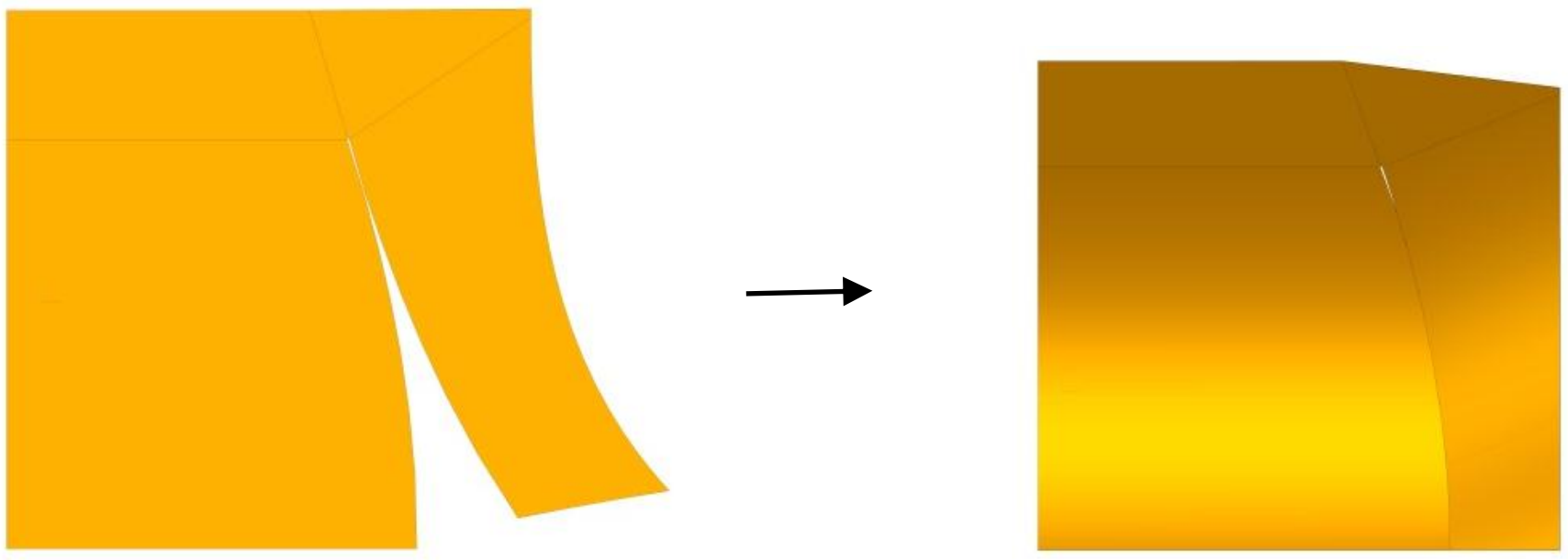

Fig 4: Planar profile before and fully-formed geometry after rolling and bending

The view on the left side of the arrow in Fig. 4 is of the geometry in its flat state. The view to the right of the arrow in Fig. 4 is the geometry formed to its 3-dimensional state after the rolling and bending operations are completed. The implementation of the sinusoidal equations using calculated inputs for the equation parameters enable the precision modeling of this multivariate geometry with a 3-D modeling program. An interesting point to note here is that the sinusoidal curve equations not only enable a precise matching of the interior edges between the primary and secondary cylinder sections when the flat state is rolled and formed to its completed 3-D state, but another set of sinusoidal equations also enable the outer and bottom edges of the secondary cylinder section to conform perfectly to the horizontal and vertical reference planes of the primary cylinder (also termed as the orthogonal planes) once the geometry has been completely formed to the 3-D state. Hence, the orthogonal characteristic of the geometry. Thus, two forms of sinusoidal curve equations are necessary for modeling the geometry defined herein: 1) One equation for the intersection of two cylinders that will be used to create the interior cuts between the primary and secondary cylinders in the flat state, and 2) a second equation for the intersection of a plane and a cylinder that will be used to create the external cuts on the secondary cylinder outer edges. The author will now provide a background of these equations to help the reader gain an insight into their applicability to this multivariate cylinder intersection geometry.

The equation required to create the internal cuts between the primary and secondary cylinder sections in the flat state was derived by the author by combining the Apostol and Mnatsakanian equations Eq. (A) and Eq. (B) referenced at the end of this paper, one as a function of the other. The result is an equation for a curve created by the intersection of two cylinders of equal radii that has then been rotated about the viewing plane:

$$y_u = \sqrt{r^2 - \left\{ r * \sin\left(\frac{x + r\alpha}{r}\right) \right\}^2} * (\sec\beta + \tan\beta) - r * \tan\beta - A$$

Eq. (1)

Where:

y = the dependent variable of the sinusoidal curve equation
u = the curve in its unwrapped state
r = the cylinder radius
x = the independent variable of the sinusoidal curve equation
α = the angle of the rotation of the curve about the viewing plane
ß = the resultant angle of cylinder intersection due to the compound rotation of the secondary cylinder section through angles θ and ω
A = a parameter which must be calculated to locate the origin of the sinusoidal curves at the origin of the coordinate system from which they will generate the interior cuts in the plane state

The parameter 'A' in the equation above is a calculation that deserves some special attention of its own, as the sinusoidal curves for creating the interior cuts between the primary and secondary cylindrical surfaces will not have their origins placed at the coordinate system from which they are based unless the position of these internal curves is adjusted with this parameter. The calculation of all the parameters required for the sinusoidal curve equations will be addressed following this introduction of the main equations.

One note here regarding Eq. (1) is that Apostol and Mnatsakanian presented two equations in their research paper: One equation to map out a curve created by the intersection of two cylinders of equal radius which has been unwrapped onto the viewing plane (i.e., Eq. A referenced in the appendix) and another equation to map out any general curve which has been rotated about the axis of the primary cylinder, or viewing plane, and unwrapped onto the viewing plane (i.e., Eq. B referenced in the appendix). The author combined Equations A and B (one as a function of the other) to create a solution for the geometry presented herein.

The second sinusoidal equation required to create the flat state of the geometry is an equation for a curve created by the intersection of a cylinder and a plane that has then been rotated about the viewing plane. This equation was derived by the author using the Apostol and Mnatsakanian equations Eq. C and Eq. B referenced at the end of this paper and is:

$$y_u = -\tan\beta * \left( r * \sin\left(\frac{x}{r}\right) * \cos\alpha + \sqrt{r^2 - \left\{ r * \sin\left(\frac{x}{r}\right) \right\}^2} * \sin\alpha \right) + (+B, -C)$$

Eq. (2)

Where:

y = the dependent variable of the sinusoidal curve equation
u = the curve in its unwrapped state
ß = the angle of intersection of the plane with respect to the primary cylinder

r = the cylinder radius
x = the independent variable of the sinusoidal curve equation
tan ß = the tangent of the angle of the cutting plane with respect to the primary cylinder axis
α = the angle of rotation of the curve about the viewing plane
+B or –C = parameters calculated to place the origin of the secondary cylinder surface external cutting curves at the desired locations so that the outer edges of the wing will align precisely with the boundaries of the orthogonal planes when the geometry is in its fully-formed 3-D state.

Eq. 2, using inputs determined by the trigonometric calculations, produces the curves that are used to create the external cuts on the outer and bottom edges of the secondary cylinder section or 'wing' as it will be referred to in this research paper. The calculations of the parameters '+B' and '–C' deserve some special attention of their own, as the sinusoidal curves used to create the exterior wing cuts would not be located properly for trimming the wing section edges to align precisely to the predetermined locations of the orthogonal boundary planes without these parameters. The calculation of all parameters required of the interior and exterior cut curve equations will be addressed following the introduction of the sinusoidal equations.

Referring again to Eq. (1), it should be noted that the curve created by this equation must be reflected across a mirror line to produce a second curve which will enable a perfect match between the interior edges that represent the cylinder intersection line when the geometry is completely formed to its 3-D state. This mirror line is calculated by differentiating Eq. (1) with respect to 'x' and then setting x equal to zero. This in effect determines the slope of both interior curves at the point x = 0 which is the origin of both curves. The resultant equation used to calculate the slope of the mirror line with respect to the local coordinate system is:

$$y' = (\sec\beta + \tan\beta) * (-\sin\alpha) \quad [\text{at } x = 0]$$

Eq. (3)

Where:

y′ = the differential of Eq. 1 with 'x' set equal to 0
ß = the calculated angle of intersection between the primary and secondary cylinder sections of the geometry
α = the calculated angle of rotation of the curve represented by Eq. (1) about the primary cylinder axis

To gain more insight into the derivation of the foregoing curve unwrapping equations, please refer to the work of Tom Apostol and Mamikon Mnatsakanian in their paper titled Unwrapping Curves From Cylinders and Cones which is available online. The equations from that paper that were used to derive the equations for application to the geometry in this research paper are referenced at the end of this paper as Equations A, B & C.

As previously noted, calculations are needed to produce the parameters A, B and C for locating the origins of the internal and external cutting curves on the geometry of the flat state of the geometry. These calculations follow:

The parameter 'A' which is required to locate the origin of both internal cutting curves on the flat state of the geometry at the origin of the coordinate systems from which those curves will generate the internal cuts is calculated as follows. When we set x equal to zero in Eq. (1), we calculate the y-value for that point:

$$y_{x=0} = \sqrt{r^2 - \{r * \sin(\alpha)\}^2} * (\sec\beta + \tan\beta) - r * \tan\beta$$

If we desire 'y' to equal zero so as to place the origin of the curve in Eq. (1) at the origin of the coordinate system from which it was generated, then we must subtract the parameter 'A' as follows (see Fig 5):

$$A = y_{x=0} = \sqrt{r^2 - \{r * \sin(\alpha)\}^2} * (\sec\beta + \tan\beta) - r * \tan\beta$$

Eq. (4)

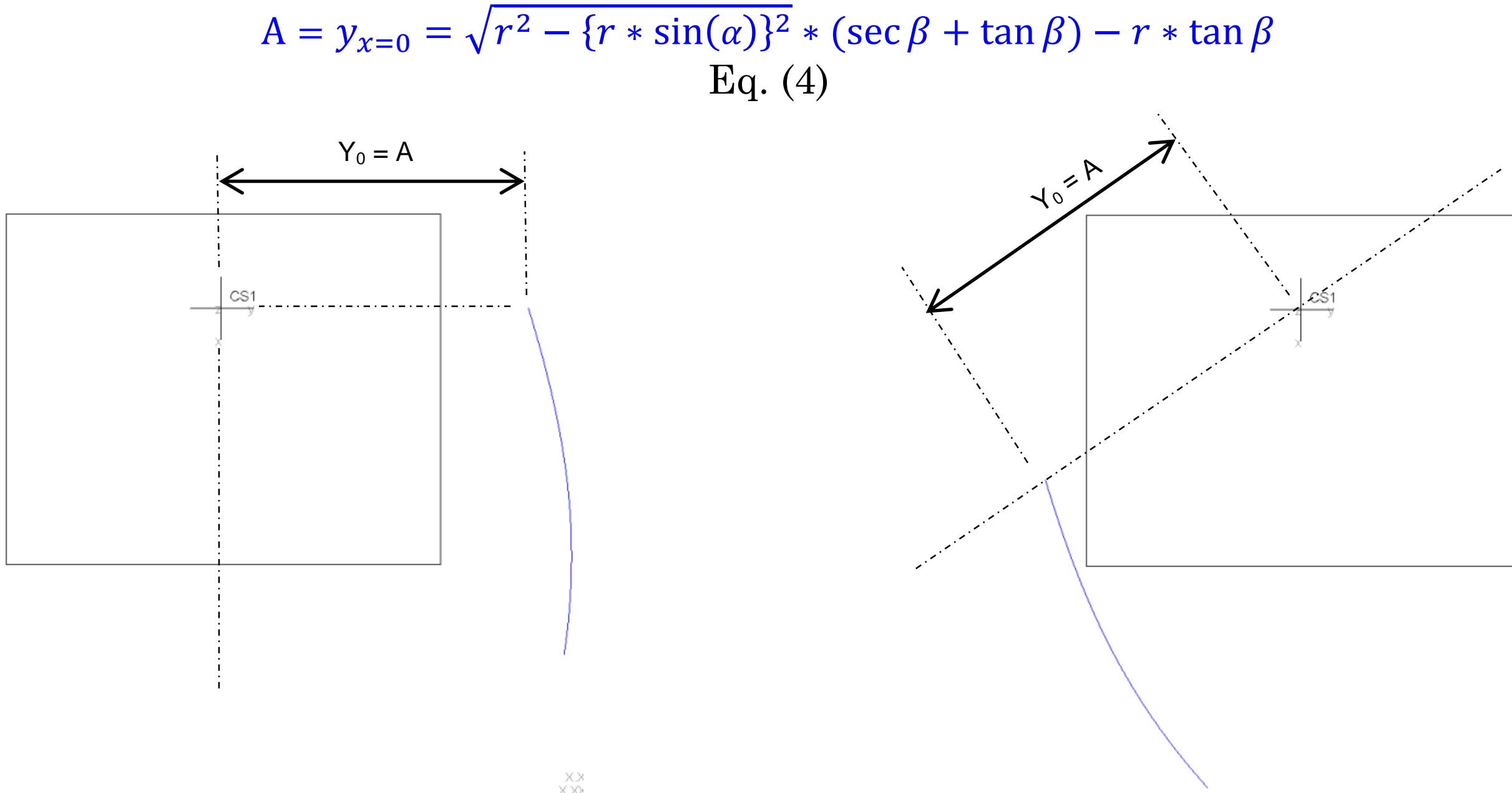

Fig. 5: 'A' = Offset for inner curve (LH view) = offset for outer curve (RH view)

A parameter 'B' is required to locate the origin of the wing end cutting curve in the correct location relative to the coordinate system from which it will be generated so that the outer edge of the wing will be bounded by (and lie entirely within) a plane that predefines the desired width of the cylinder intersection geometry in its 3-D state and which is orthogonal to the axis of the primary cylinder. To this end, we set:

a = X – I, where:

X = the preselected overall width of the cylinder surface geometry in its 3-D state;

I = the distance between the left vertical edge of the primary cylinder surface geometry and the predetermined location of the apex of the internal cut curves (i.e., the location of the local coordinate system of the internal cut curves). Then,

c = a / sin ß = wing offset component (i.e., wing width) in the flat state;

where 'c' is the width of the wing in its flat state. This is the offset required to place the origin of the wing end cut curve in the precise location on the flat blank so that the desired overall width of the geometry will be achieved following the rolling and bending operations to form the geometry to the 3-D state. A visual depiction of 'c' can be seen in Fig. 7 as the straight line segment located at the top of the wing.

We next set x = 0 in Eq. (2) so that:

$$y_{x=0} = -\tan\beta * \left(\sqrt{r^2} * \sin\alpha\right)$$

And let:

$$d = y_{x=0} = -\tan\beta * \left(\sqrt{r^2} * \sin\alpha\right) = \text{zero offset component}$$

This value of 'd' is the offset of the origin of the wing end cut curve from the local coordinate system from which it was created when 'x' was set to zero. This offset, similar to the other sinusoidal curve offsets, is required due to the curve being rotated by an angular amount 'φ' about the primary cylinder axis in the viewing plane. To calculate the parameter 'B' required in Eq. (2) to offset the wing end cut correctly, one must add the amount of offset yielded by Eq. (2) when x is equal to zero (i.e., 'd', or the amount of offset that would be required to locate the origin of the wing end cut curve at the origin of its local coordinate system in the viewing plane) with the offset value 'c' which will set the wing width in the flat state to a value which will result in the predetermined overall width of the geometry when the geometry is in its formed state. Thus,

$$B = c + d$$

Eq. (5)

Please see Figs. 6 and 7 for a visual depiction of the offsets required for the location of the curve origin for the wing end cut.

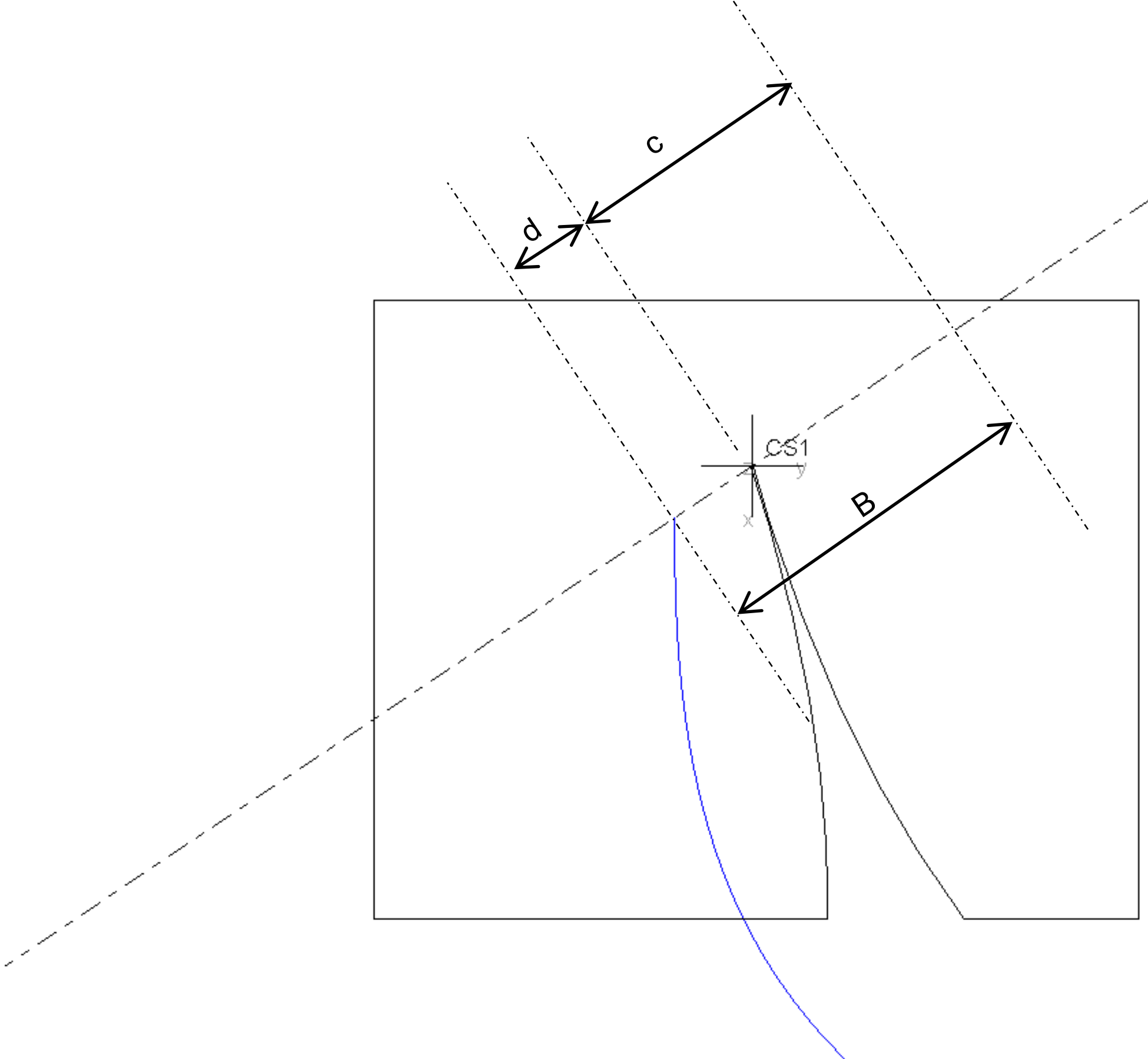

Fig. 6: Offset 'B = c + d' for the wing end cut

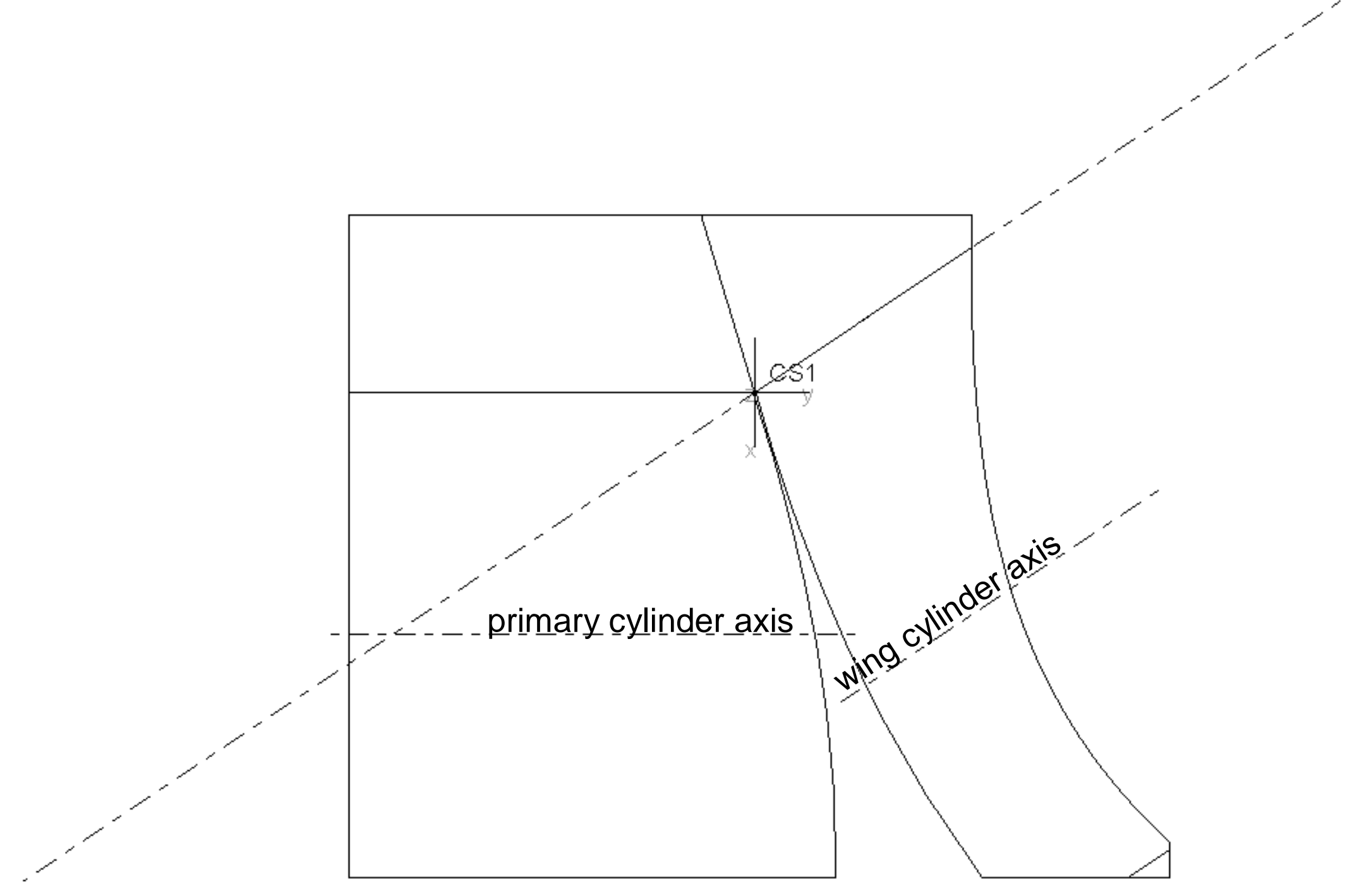

Fig. 7: Wind end cut completed in the flat state

A parameter 'C' is required to locate the origin of the wing bottom cutting curve so that the bottom edge of the wing, once the geometry is formed to the 3-D state, will lie in the same horizontal plane as the lower edge of the primary cylinder surface section following the rolling and bending operations. This horizontal plane is orthogonal to the primary cylinder axis and the vertical plane which bounds the outer edge of the wing when the geometry has been fully formed to its 3-D state. The parameter 'C' is calculated as follows:

Calculate the arc length of the primary cylinder surface section by measuring height of the primary cylinder surface in its flat state minus the height of the flat upper portion. This value should have already been predetermined by the design of the overall dimensions of the flat blank and by the chosen location of the coordinate system from which the internal cutting curves would be generated.

Using Eq. (1), calculate the y dimension for the value of x (arc length) measured above. This (x,y) point is in effect the end point of the interior curve on the primary cylinder surface relative to the coordinate system from which that internal cutting curve was generated (i.e., the primary coordinate system CS1).

Since we will be working in the transformed (rotated) coordinate system of the wing section (i.e., CS2, or the secondary cylinder surface coordinate system), x and y in the primary coordinate system CS1 become x' and -y' in the wing secondary coordinate system CS2. This (x',-y') point is in effect the end point of the interior cut curve on the wing relative to the rotated (transformed) coordinate system CS2.

Now that we have values for the abscissa x' and ordinate -y' for the end point of the wing internal cut curve, we can enter these values of x' and –y' into Eq. (2) to calculate the amount of offset 'C' which will be required to locate the wing bottom cut curve in the proper location; i.e., in the same horizontal orthogonal plane as the bottom edge of the primary cylinder surface once the geometry has been fully formed to the 3-D state:

$$y_u = -\tan\beta * \left( r * \sin\left(\frac{x}{r}\right) * \cos\alpha + \sqrt{r^2 - \left\{ r * \sin\left(\frac{x}{r}\right) \right\}^2} * \sin\alpha \right) - C$$

$$\xrightarrow{yields} \quad C = -y_u - \tan\beta * \left( r * \sin\left(\frac{x}{r}\right) * \cos\alpha + \sqrt{r^2 - \left\{ r * \sin\left(\frac{x}{r}\right) \right\}^2} * \sin\alpha \right)$$

Eq. (6)

This is the calculation to determine the parameter 'C' which will be required in Eq. (2) to locate the wing bottom cutting curve at such a location that the end points of the interior and exterior curves will align precisely once all cuts have been made and the geometry is formed into its 3-D state. Please see Figs. 8, 9 and 10 for a visual depiction of the offsets required for the location of the curve for the wing bottom cut.

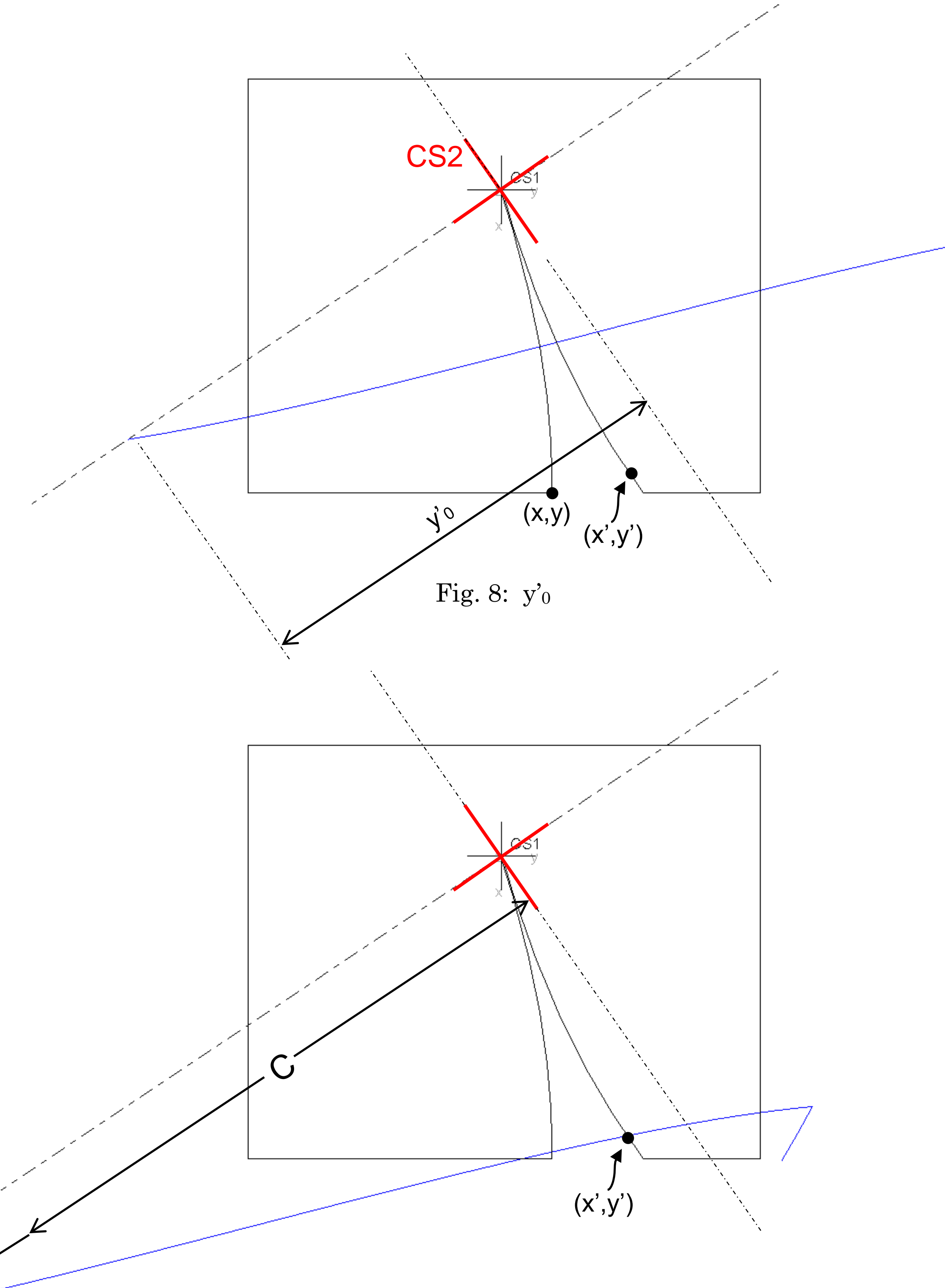

Fig. 8: $y'_0$

Fig. 9: C = f(x’,-y’)

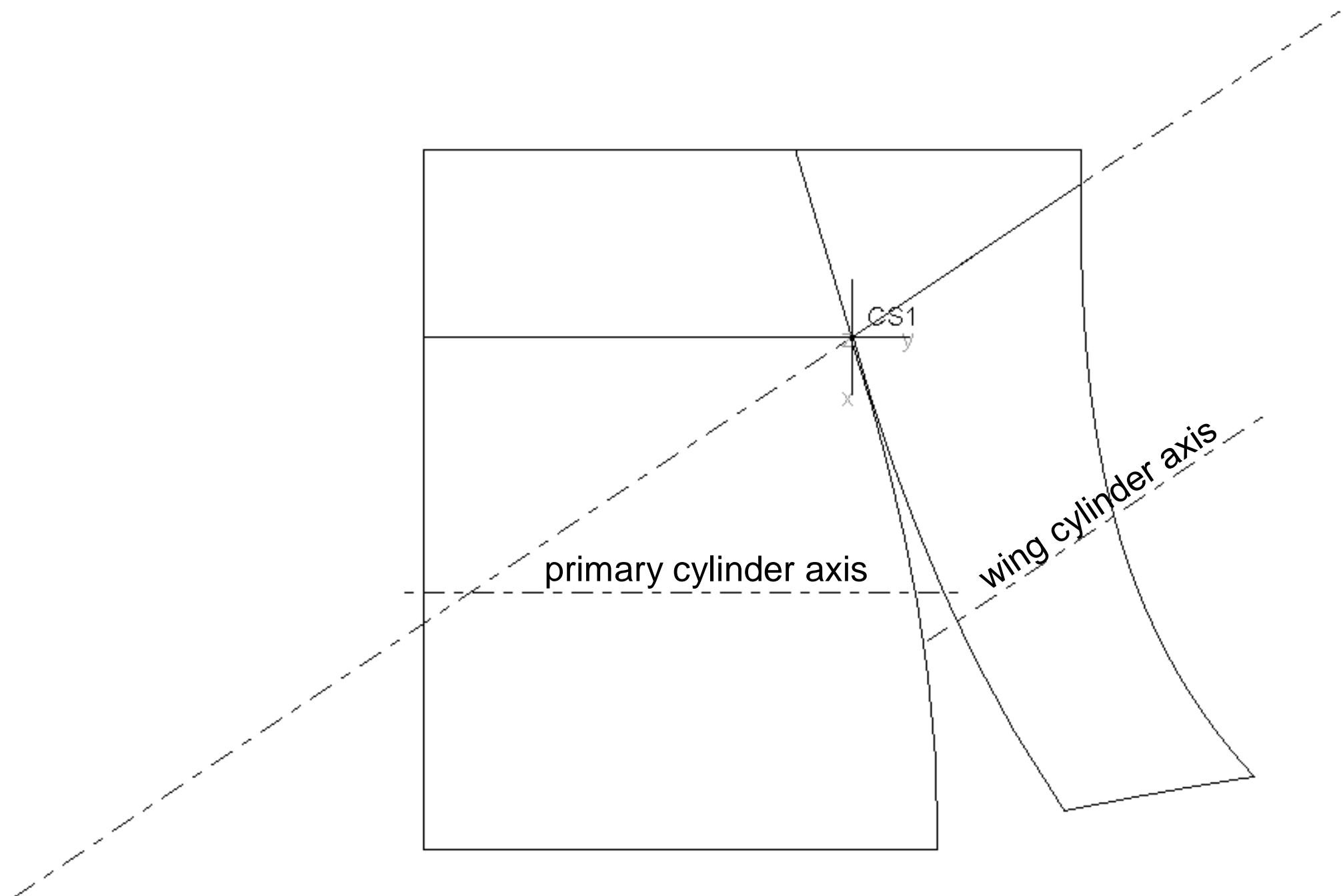

Fig. 10: Bottom wing cut in the flat state completed

Please note that the parameters A, B and C that have now been defined cannot be determined until the values of $\alpha$ and $\beta$ are calculated; and $\alpha$ and $\beta$ are, in turn, determined by the choices of the primary and secondary angles of rotation of the secondary cylinder surface (or wing) with respect to the primary cylinder surface.

A visual depiction of the sinusoidal cutting curves created as derivations of the Apostol and Mnatsakanian curve unwrapping equations and as applied to the geometry presented herein can be seen in Fig. 11. An interesting point to make here is that the sinusoidal curves required at the end and bottom of the wing in the plane state of the geometry result in curved edges that lie completely within horizontal and vertical planes that are orthogonal to the primary cylinder axis once the wing has been rolled and formed into its finished 3-D state.

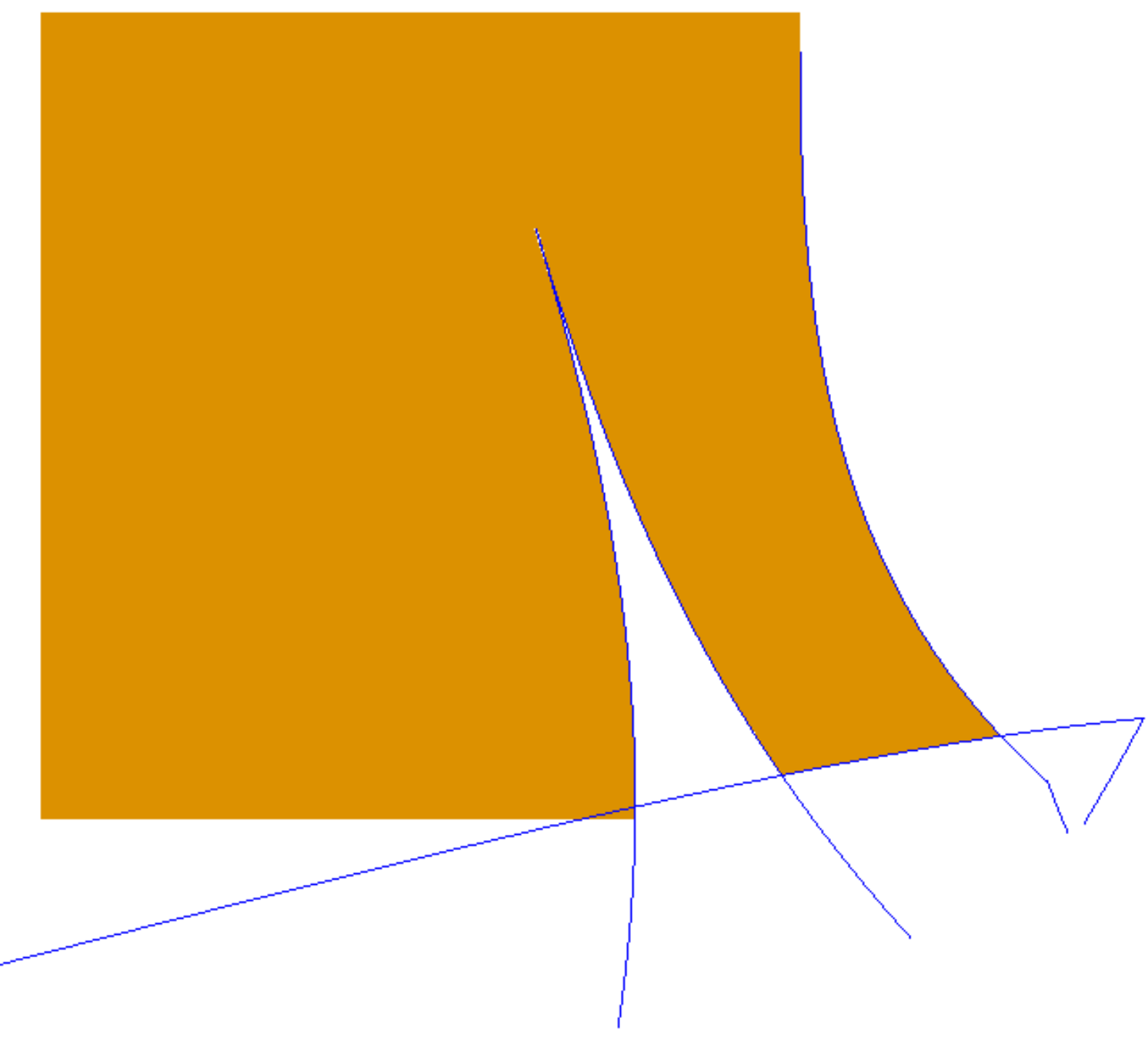

Fig. 11: Sinusoidal curves define the profile of the geometry in its flat state

The four blue curves illustrated on the flat-state geometry in Fig. 11 that extend beyond the boundaries of the geometry are derivations of the Apostol and Mnatsakanian equations which are used to define the flat state profile of the geometry. Although not intuitive from the flat-state view, the curves of the interior and exterior cuts created on the secondary cylindrical surface (which will alternately be referred to as 'the wing' in this paper) are not constant radii or straight line segments but are instead sinusoidal curves mapped out by derivations of the Apostol and Mnatsakanian equations using parametric inputs determined by the author's trigonometric calculations. The curves are overlaid on the blank in its flat state, relative to their local coordinate systems, to create the profile shown in Fig. 11 prior to forming the final 3-D shape of the geometry. It will be noted here that other parameters will be required to enable the transformation of the flat state to the fully-formed 3-D geometry via rolling and bending operations, and these parameters are illustrated at the top of the flat-state geometry in Fig. 12. Fig. 13 illustrates the transition of the flat state to the completely formed 3-D state via the rolling and bending operations.

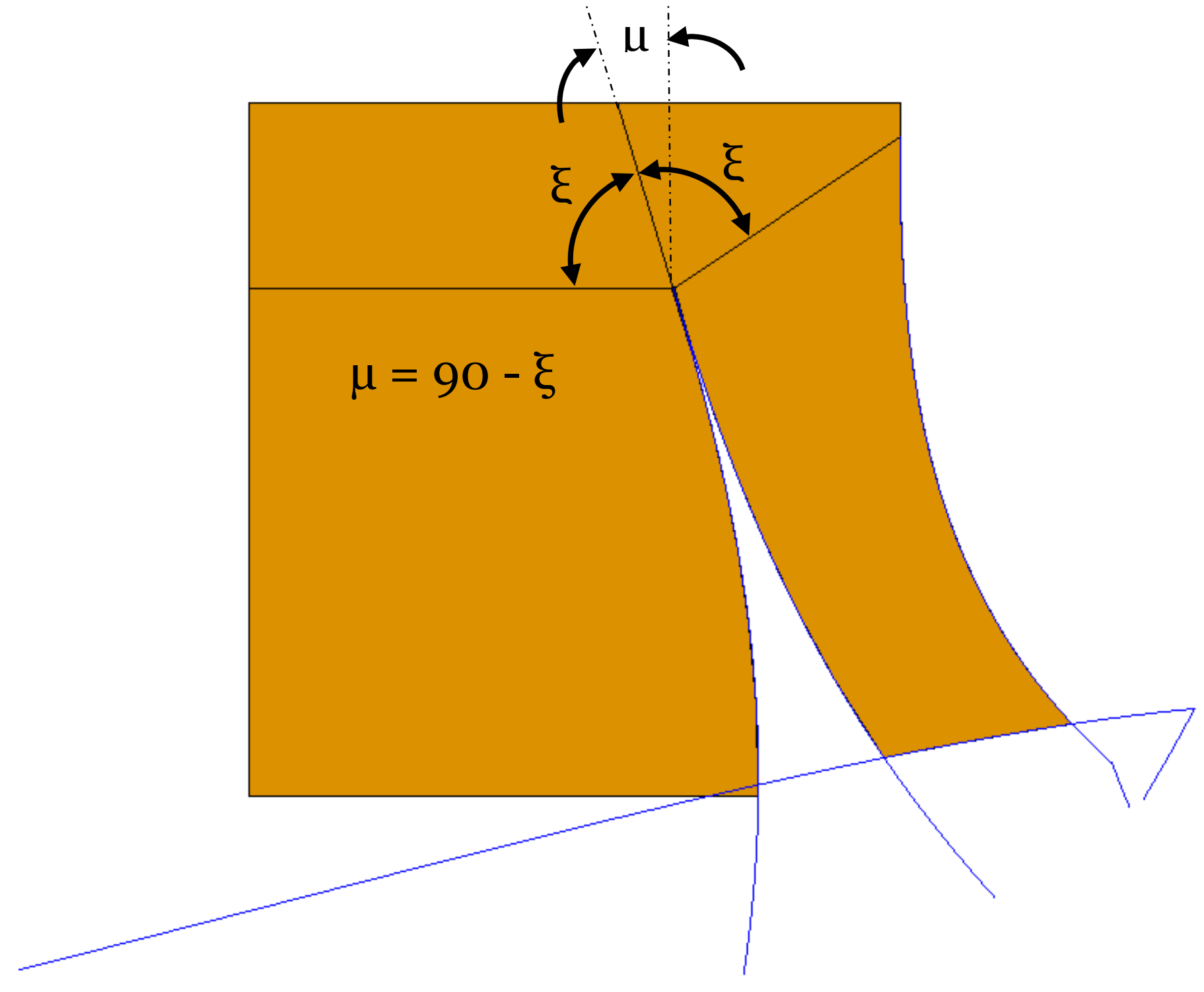

Fig. 12: μ = Angle of the split line for mirroring the interior curves

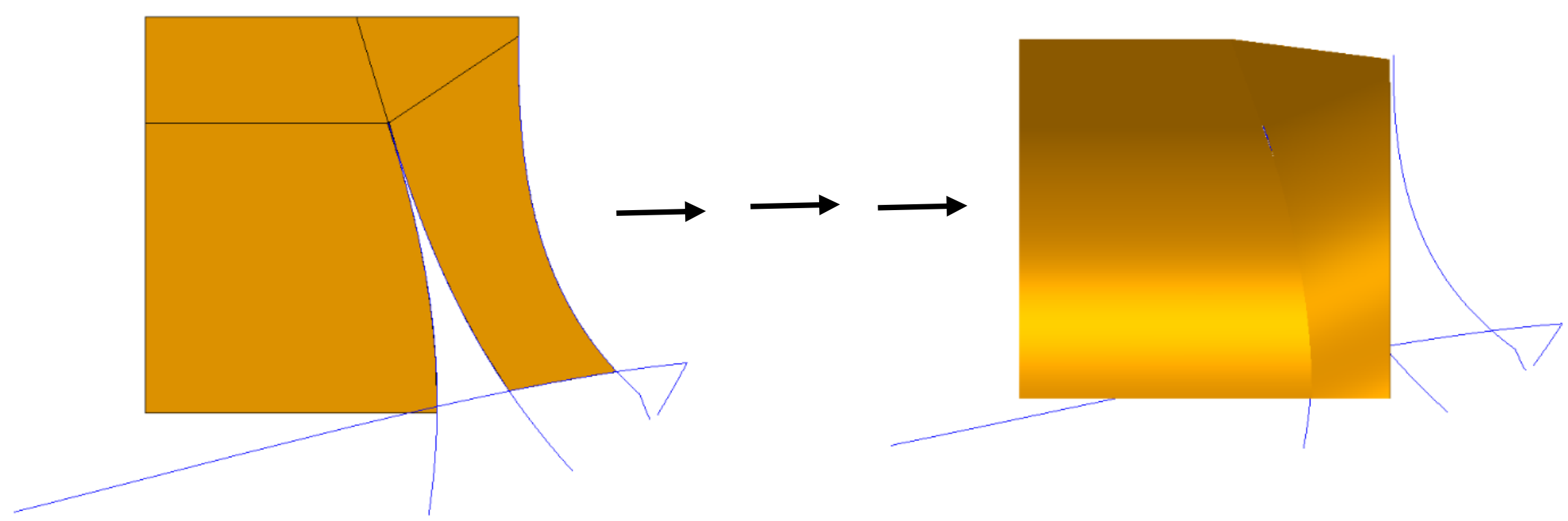

Fig. 13: Transition from the flat state to the 3-D state via the rolling and bending operations

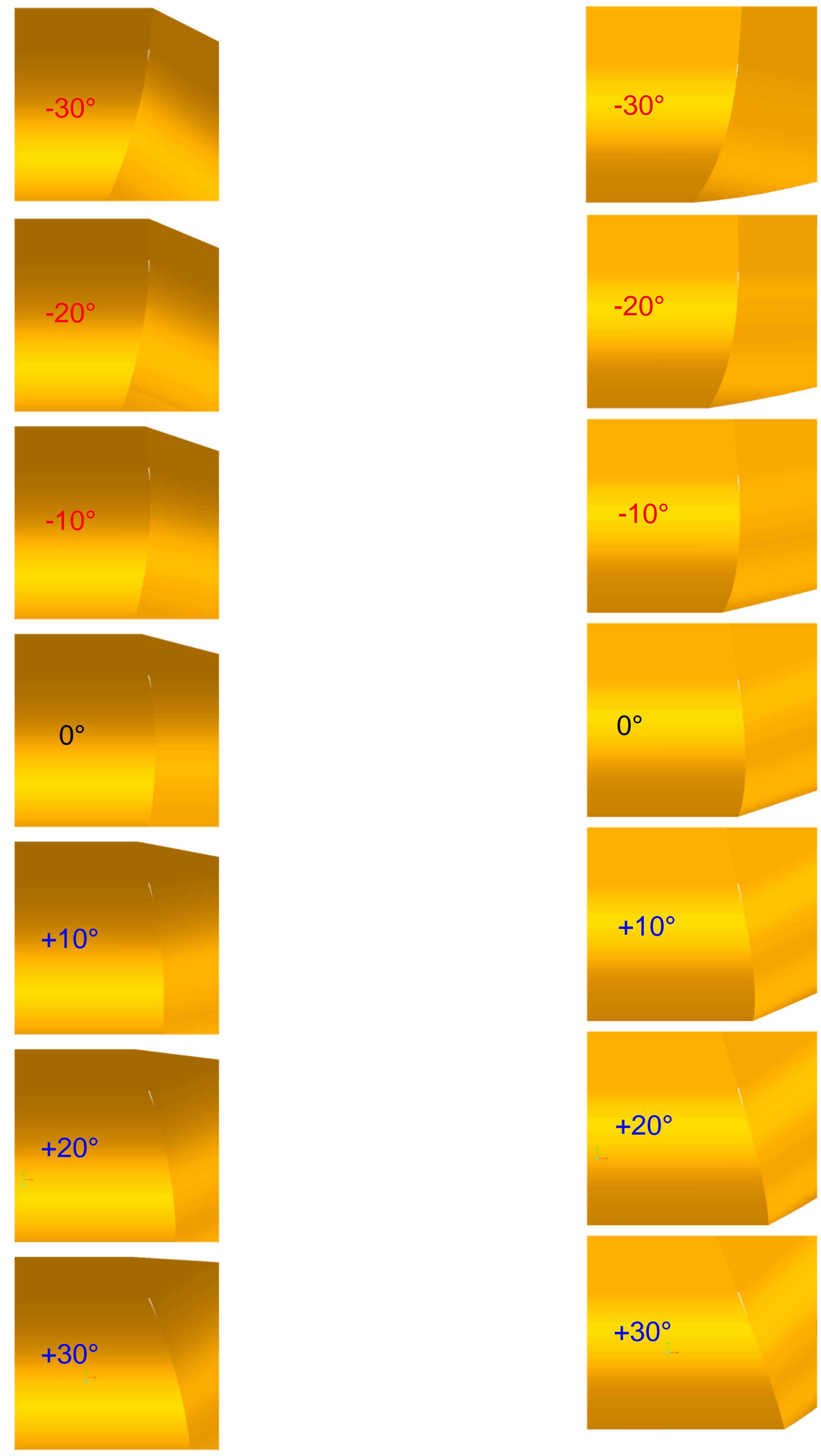

Fig. 14: Geometric variation as $\theta$ is held constant at 30° while $\omega$ is varied

Fig. 14 illustrates the variable feature of the 3-D geometry as the primary direction of wing rotation θ is held constant at 30° while the secondary direction of wing rotation ω is varied from -30° to +30° in 10° increments. For this illustration in Fig. 14, the column on the left-hand side illustrates the alignment of the viewing plane normal to the horizontal orthogonal plane, and the column on the right-hand side illustrates the alignment of the viewing plane normal to a tertiary orthogonal plane. Notice in both viewing perspectives that not only does the general shape of the wing change as ω is varied, but also the length and orientation of the intersection line between the primary and secondary (wing) cylinders. One would expect similar geometric variations if the primary direction of wing rotation θ were varied while the secondary direction of wing rotation ω is held constant.

Illustrated in Fig. 15 are snapshots of the interesting geometry formed when the variable instances of Fig. 14 are overlaid via their common coordinate system:

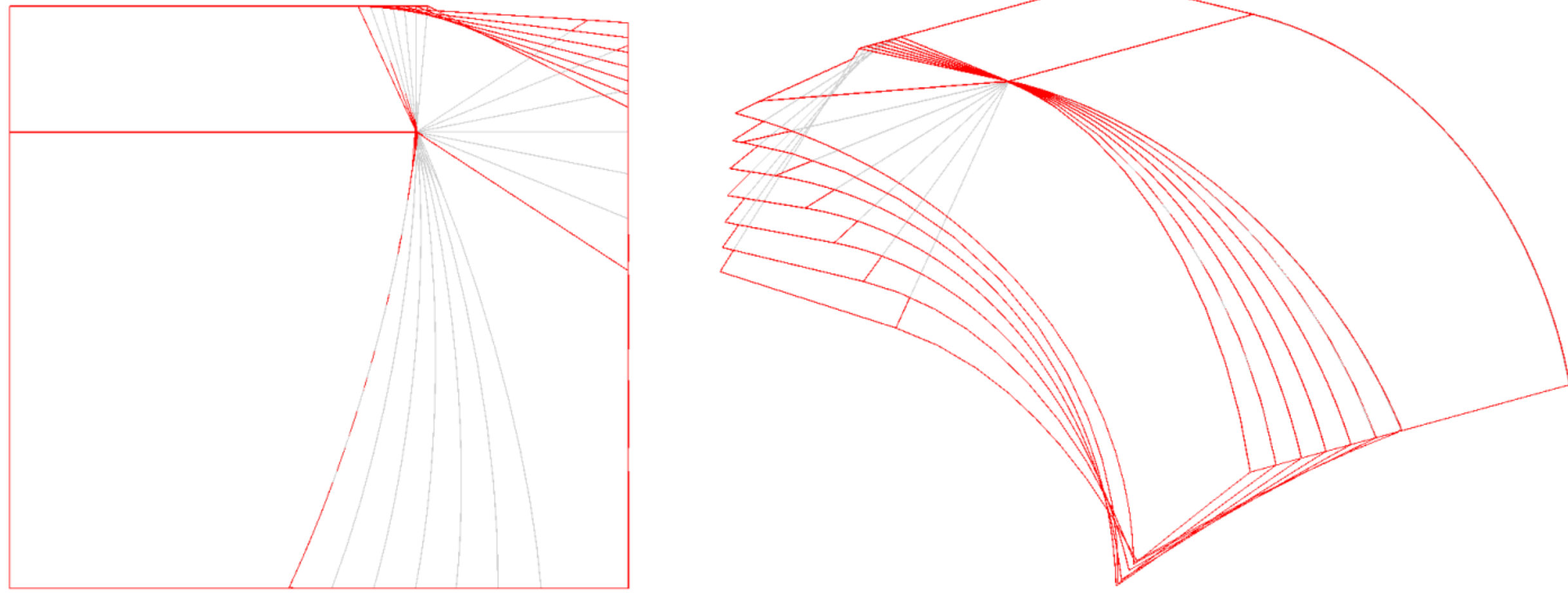

Fig. 15: View of the overlay of the geometric variations of the geometry as ω is varied while θ is held constant

We'll now review the source of the parameters which provide the variable inputs required of the sinusoidal curve equations to yield the precise outline of the multivariate cylinder intersection geometry in its flat state. Fig. 16 illustrates the set of intersecting cylinder axes formed as a result of the choices of wing angle θ and wing rotation ω. For the purposes of this discussion, 'wing angle' will refer to the wing rotation in the primary direction θ while 'wing rotation' will refer to the rotation of the wing in the secondary direction ω.

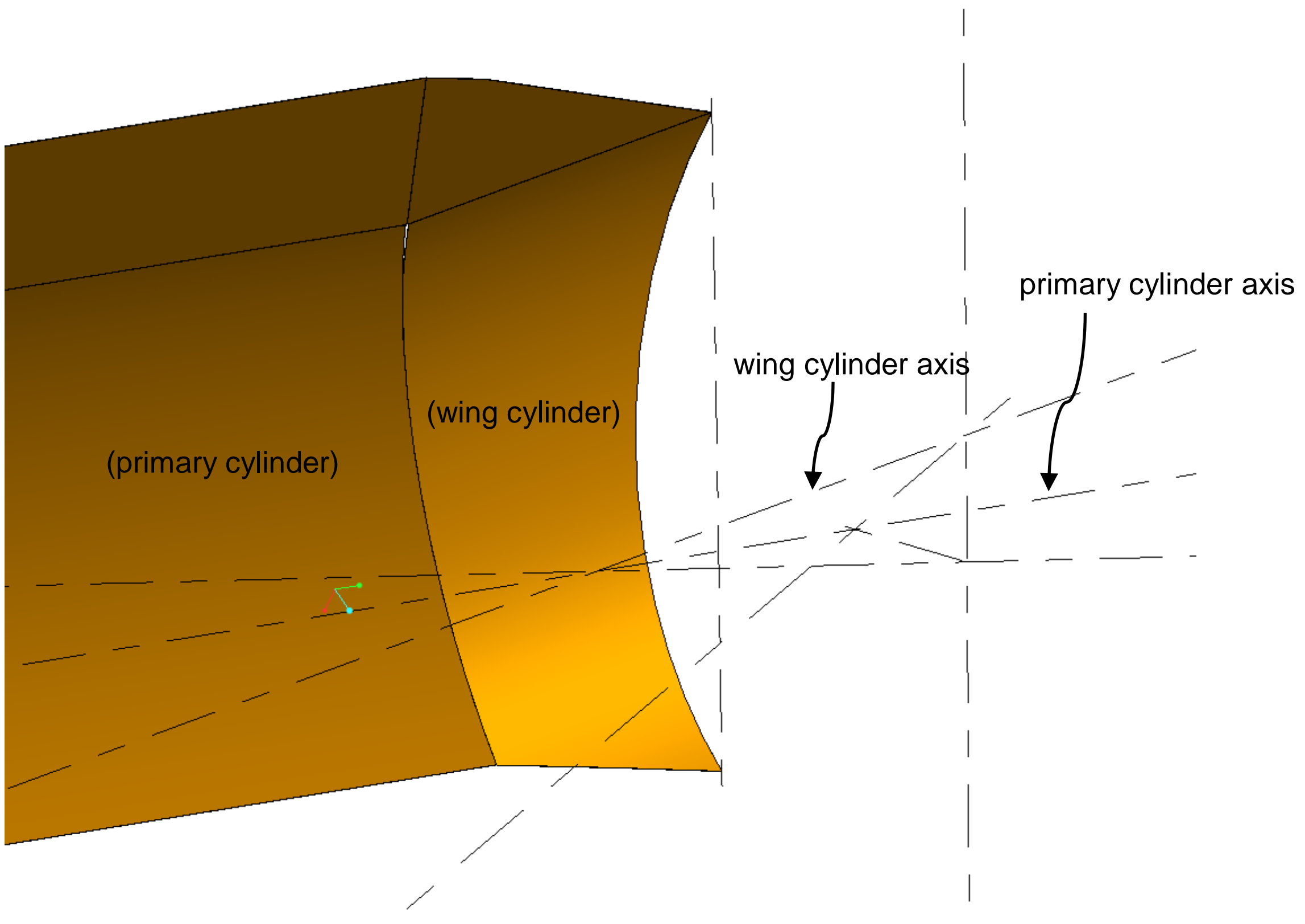

Fig. 16: Intersection of the primary cylinder and secondary cylinder axes as a result of the choices of wing angle θ and wing rotation ω

Challenges specific to this cylinder intersection geometry problem which were not addressed explicitly by the curve unwrapping equations referenced at the end of this research paper are 1) the case of a curve created by the intersection of two cylinders that is subsequently rotated about the axis of the primary cylinder, and 2) the case of a curve created by the intersection of a plane and cylinder that is subsequently rotated about the axis of the primary cylinder. Thus, the initial challenge of arriving at a mathematical solution for constructing the multivariate cylinder intersection geometry was two-fold: in the first sense, to find the resultant angle between the axes of the primary and secondary cylinders when the two directions of secondary cylinder surface rotation were chosen; and in the second sense, to find the angle between the plane formed by the axes of the primary and secondary cylinders, and the primary cylinder's horizontal plane. It will be seen that both of these angles provide parametric inputs into the sinusoidal curve equations which will yield the precise orientation of the curves on the flat state of the geometry.

After much contemplation on the concept of sinusoidal curve orientation in this multivariate geometry application and how to calculate the parameters required of the curve equations to precisely orient the cutting curves on the one-piece blank, the author discovered that if an irregular tetrahedron is constructed at the intersection of the primary and secondary cylinder axes, using the primary cylinder's horizontal and vertical datum planes (i.e., the orthogonal planes) as boundaries, then all the parameters required of the curve equations for the cylinder-to-cylinder and plane-to-cylinder intersections in this application can be extracted from this tetrahedron. Fig. 17 is a visual depiction of the tetrahedron constructed for this purpose. We will return to this tetrahedron later where it will be demonstrated how the parameters for the sinusoidal curve equations are extracted therefrom.

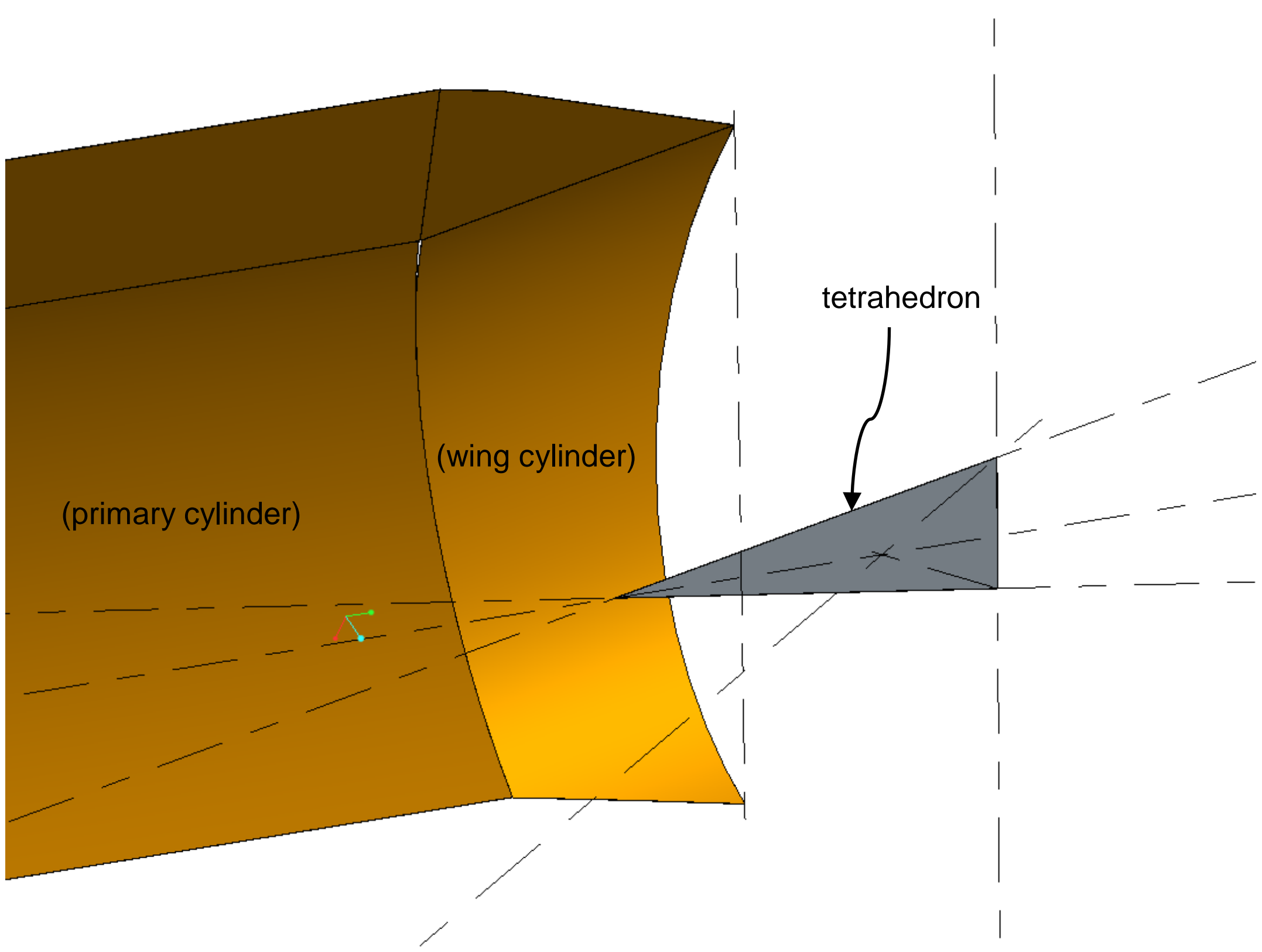

Fig. 17: Irregular tetrahedron formed by the intersection of the main and wing cylinder axes and the planes orthogonal to the primary cylinder

The first step in the construction of the multivariate cylinder surface section geometry is to calculate the resultant angle formed between the primary and secondary cylinder axes for choices of θ and ω which are the primary and secondary directions of rotation of the secondary cylinder surface (wing) with respect to the primary cylinder surface, respectively. We will designate this angle as λ. Calculating this angle is accomplished by constructing a diagram of the intersection of the primary cylinder and wing cylinder axes and then using the law of cosines in two successive applications to arrive at the solution. Please see Fig. 18 for the trigonometric calculation details and Fig. 19 for a visual depiction of the geometry:

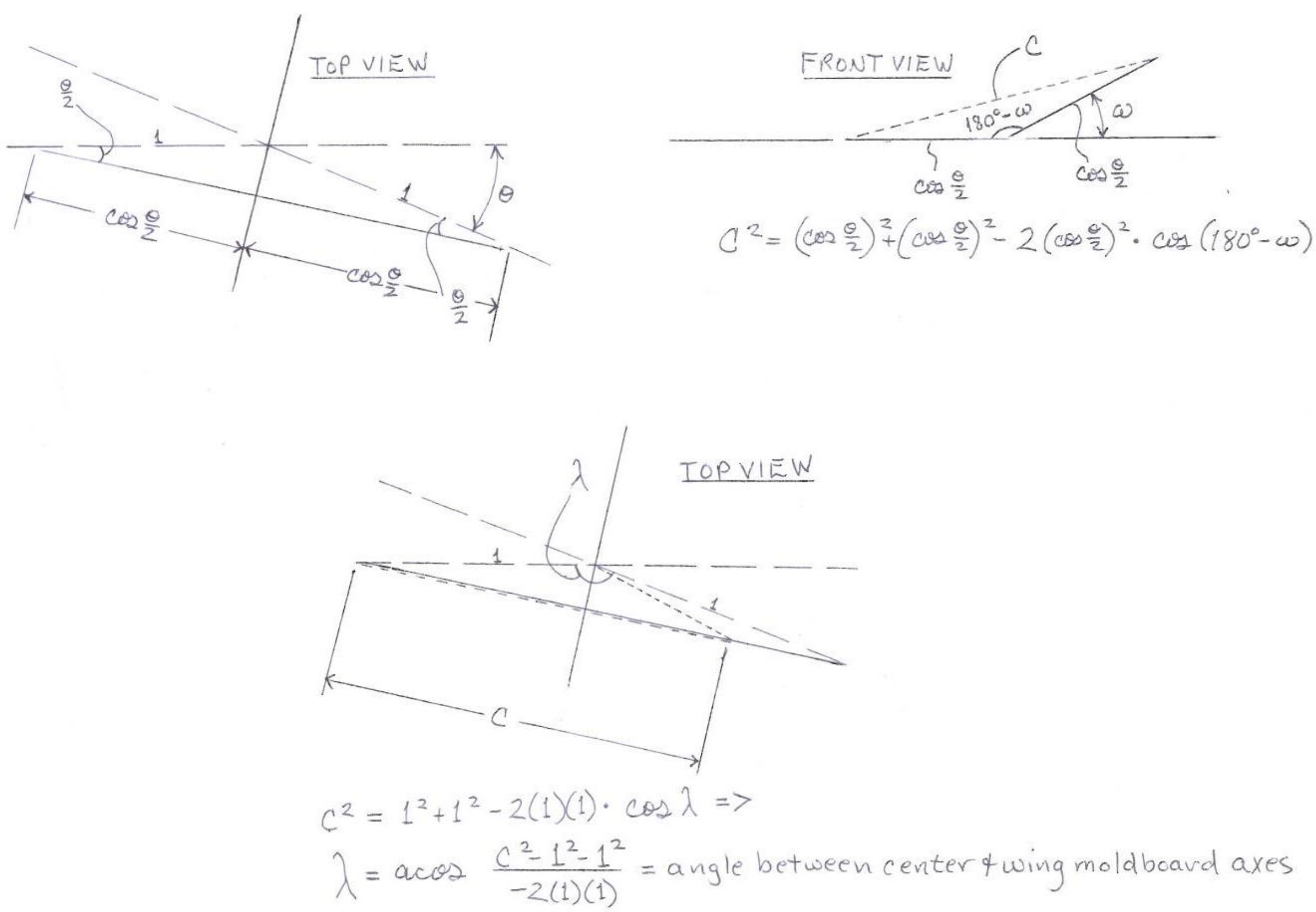

Fig. 18: Calculation of the angle formed between the axes of the primary and secondary cylinders for selected values of θ and ω

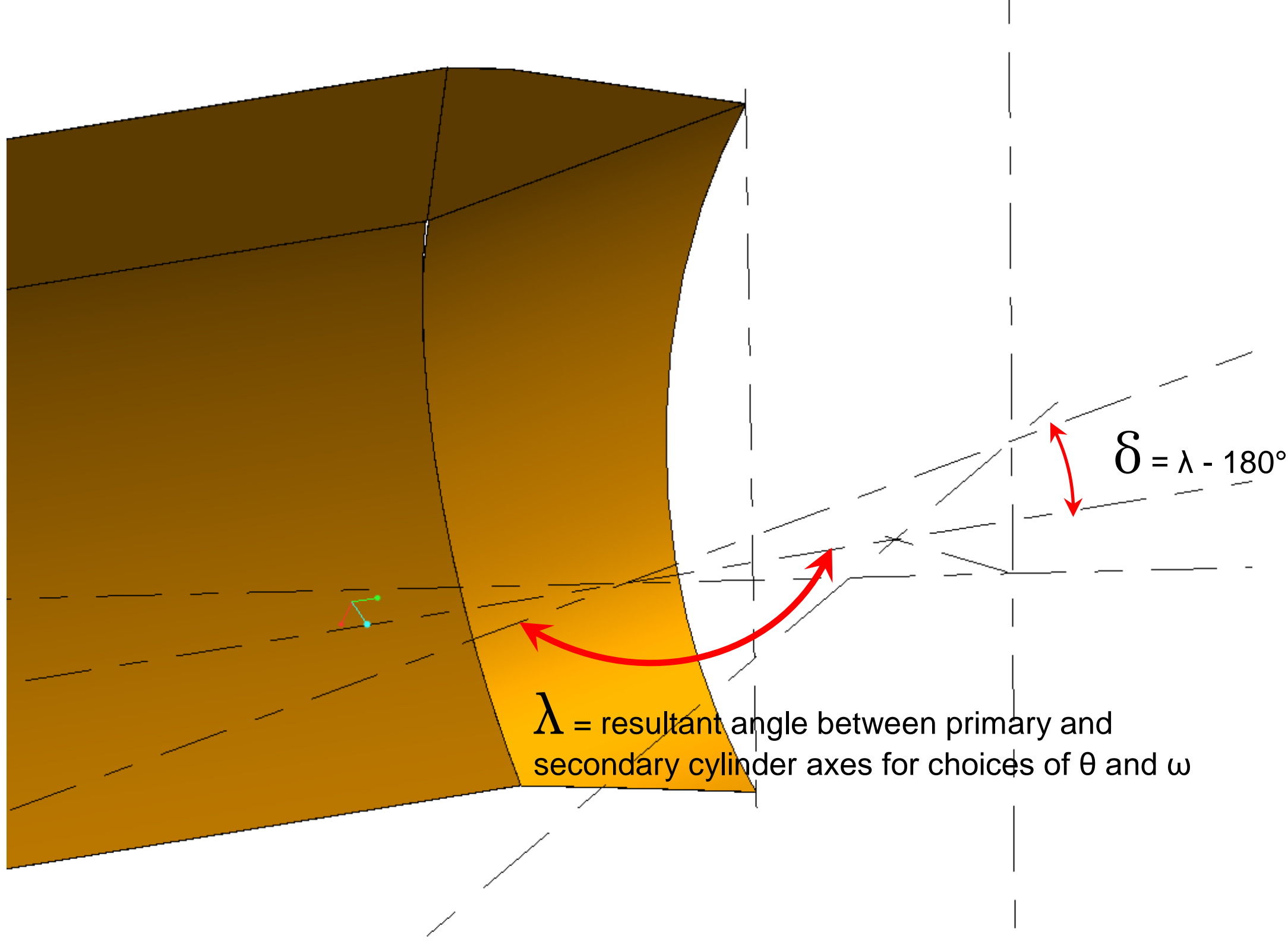

Fig. 19: λ = Resultant angle formed between the primary and secondary cylinder axes for selected rotations of the secondary cylinder surface section (wing)

Once λ is determined, then δ, or the resultant tilt of the wing axis with respect to the primary cylinder axis is calculated as λ - 180°. To illustrate a numerical example of the effect that the rotation ω of the wing about the secondary axis of rotation has on the resultant tilt of the wing axis with respect to the primary cylinder axis, consider the following: For wing rotations of 30° and 20° for θ and ω, respectively, the wing tilt angle δ is calculated to be 35.9277°. Compare this to the resultant wing tilt angle of 30° if ω (the secondary direction of rotation) were set to 0°. In other words, increasing the second degree of wing rotation from 0° to 20° increases the resultant wing tilt angle from 30° to 35.9277°.

The next step in the calculations is to determine the angle between the horizontal plane of the primary cylinder surface (also referred to as the horizontal orthogonal plane of the geometry) and the horizontal plane of the wing. We will designate the wing plane as the plane formed by the intersection of primary and secondary (wing) cylinder axes. For cylinder intersection geometry where there is only wing rotation in the primary θ direction (i.e., the wing is angled in the direction of θ but not rotated in the direction of ω), the angle between the wing plane and the horizontal plane of the primary cylinder (or horizontal orthogonal plane) is always 0°. For the example presented in this paper, the orientation of the horizontal orthogonal plane for the overall geometry is fixed with respect to the primary cylinder such that the small planar section at the top of the primary cylinder surface section is oriented at an angle of γ from the horizontal orthogonal plane. Therefore, the angle between the horizontal orthogonal plane and the flat section tangent to the top of the primary cylinder surface remains constant as θ and ω are varied to produce the multivariate geometry. A tertiary orthogonal plane is established normal to the flat section tangent to the top of the primary cylinder surface to complete the orthogonal outer boundary of the geometry. Thus, the angle between the tertiary orthogonal plane and the horizontal orthogonal plane is designated as 90°-γ. Please see Fig. 20 for a visual definition of the horizontal, vertical and tertiary orthogonal boundary planes.

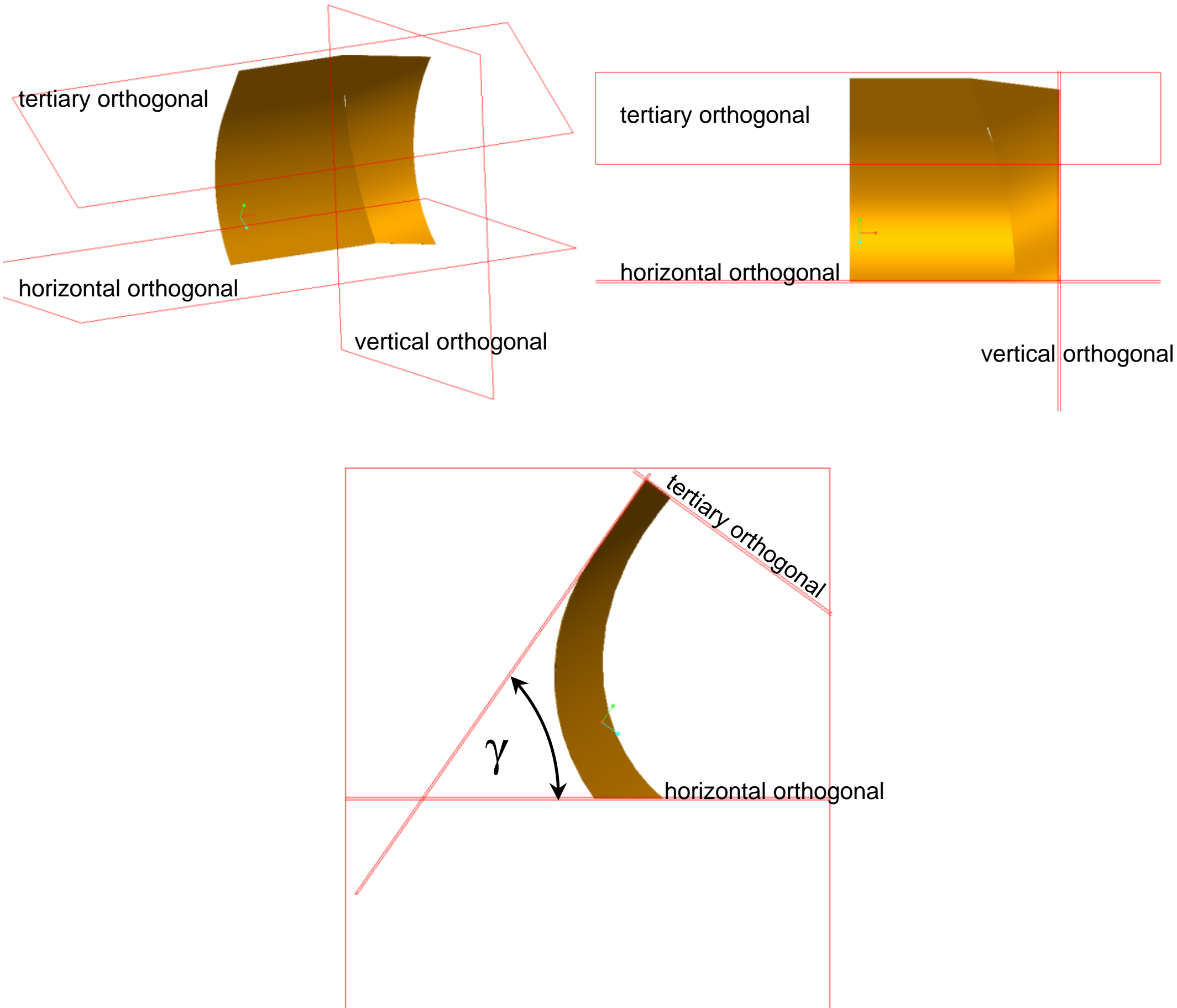

Fig. 20: Definition of the horizontal, vertical and tertiary orthogonal boundary planes of the cylinder intersection geometry

We'll now refer to Fig. 21 for a visual depiction of the angle between the primary cylinder and wing horizontal planes that results from the chosen values of wing angle and wing rotation, θ and ω, respectively. We will designate this angle as σ, or the interior angle between the horizontal orthogonal plane and the wing horizontal plane. This is an important angle which will be used in calculations to determine the orientation for the interior and exterior wing cuts. As was for the case of the calculation of the angle between the cylindrical axes of the primary and wing cylinder surfaces, there is some hand calculation required to produce this angle. Please see Fig. 22 for a diagram and the details of this calculation.

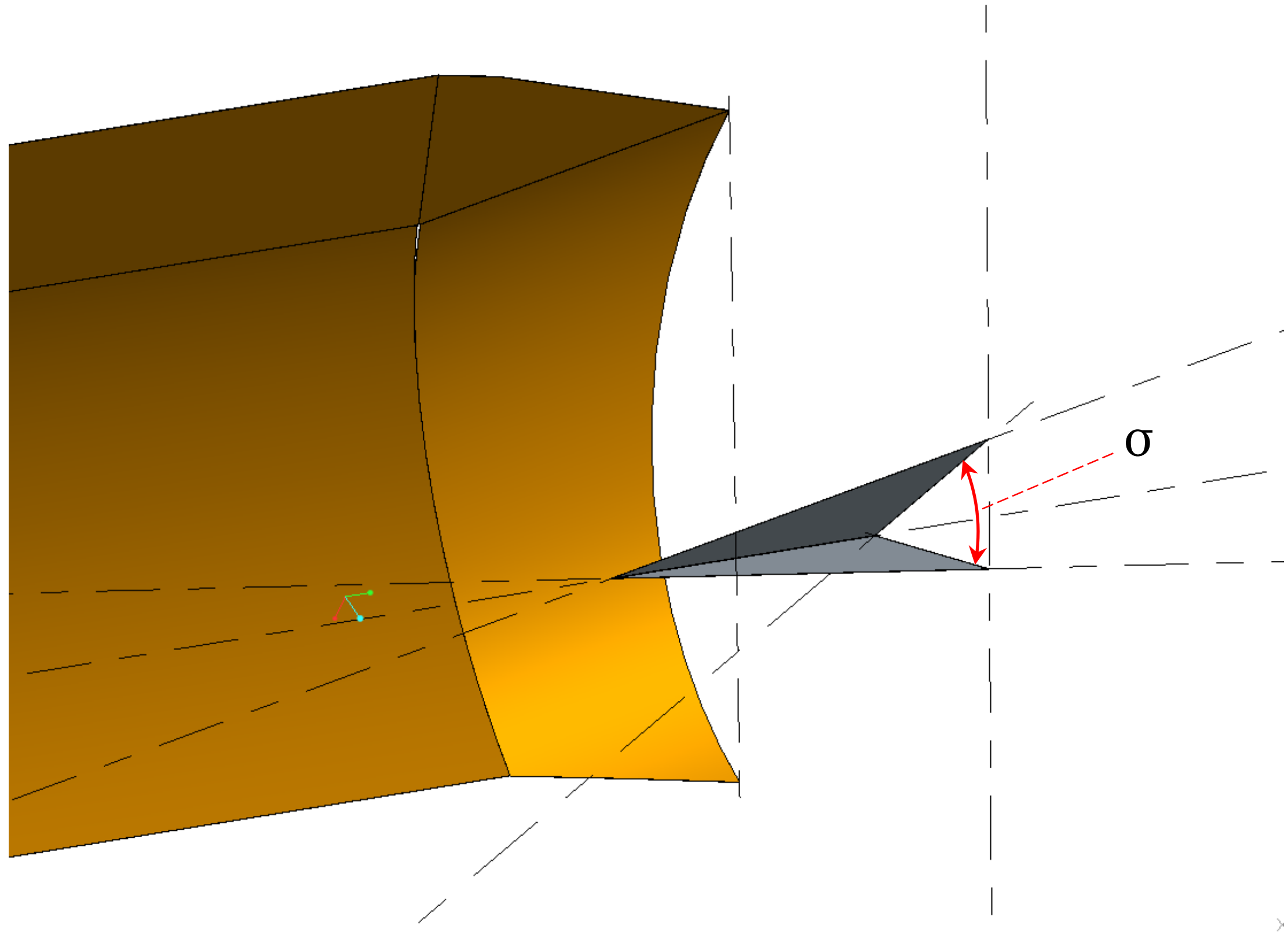

Fig. 21: σ = Resultant angle between the horizontal orthogonal and wing horizontal planes for choices of θ and ω

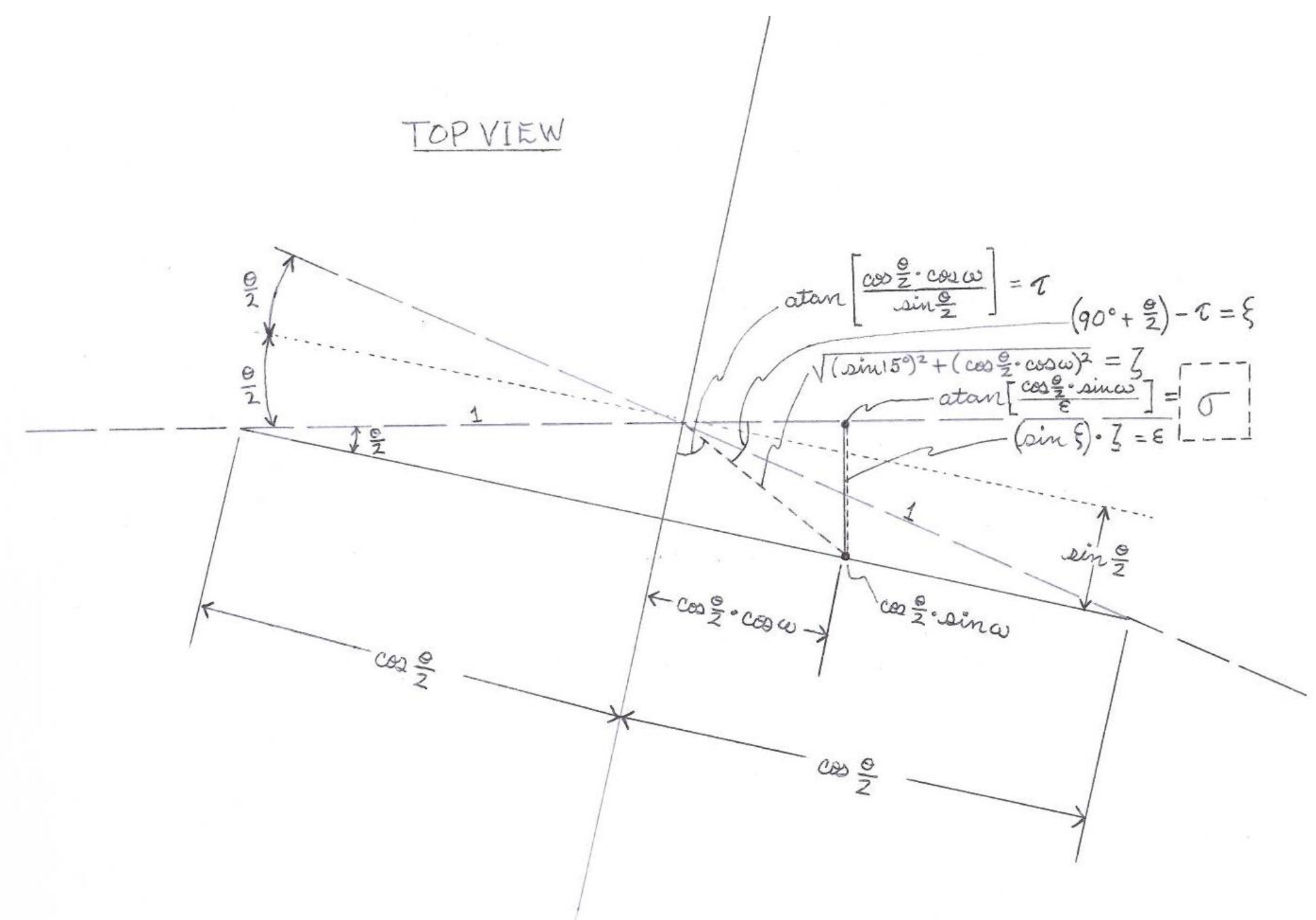

Fig. 22: Calculation of the angle between the horizontal orthogonal and wing horizontal planes

Now that the angle between the primary cylinder and wing cylindrical axes and the angle between the horizontal orthogonal and wing horizontal planes have been calculated, these parameters can be combined with other parameters extracted from the tetrahedron to write the curve equations for the interior and exterior wing cuts. All angular values extracted from the tetrahedron will of course be unique to the each selection of wing angle θ and wing rotation ω made. We will proceed with the details of these calculations.

Please refer to Fig. 23. Geometrically, ß is measured on the tetrahedron as the angle between the wing cylindrical axis and an axis formed by the intersection of the vertical orthogonal plane of the geometry and the wing horizontal plane. This is one of the two angles needed to write the equations for the curves used to create the interior cuts. In particular, this angle defines the resultant angle of intersection between the primary cylinder and wing cylinder sections in terms of the coordinate system of the Apostol and Mnatsakanian equations. The equation for ß is derived from the angle measured between the primary cylinder and wing cylinder axes determined earlier and is calculated as:

$$\delta - 90°$$

where δ is the resultant angle of tilt between the wing cylinder axis and the primary cylinder axis.

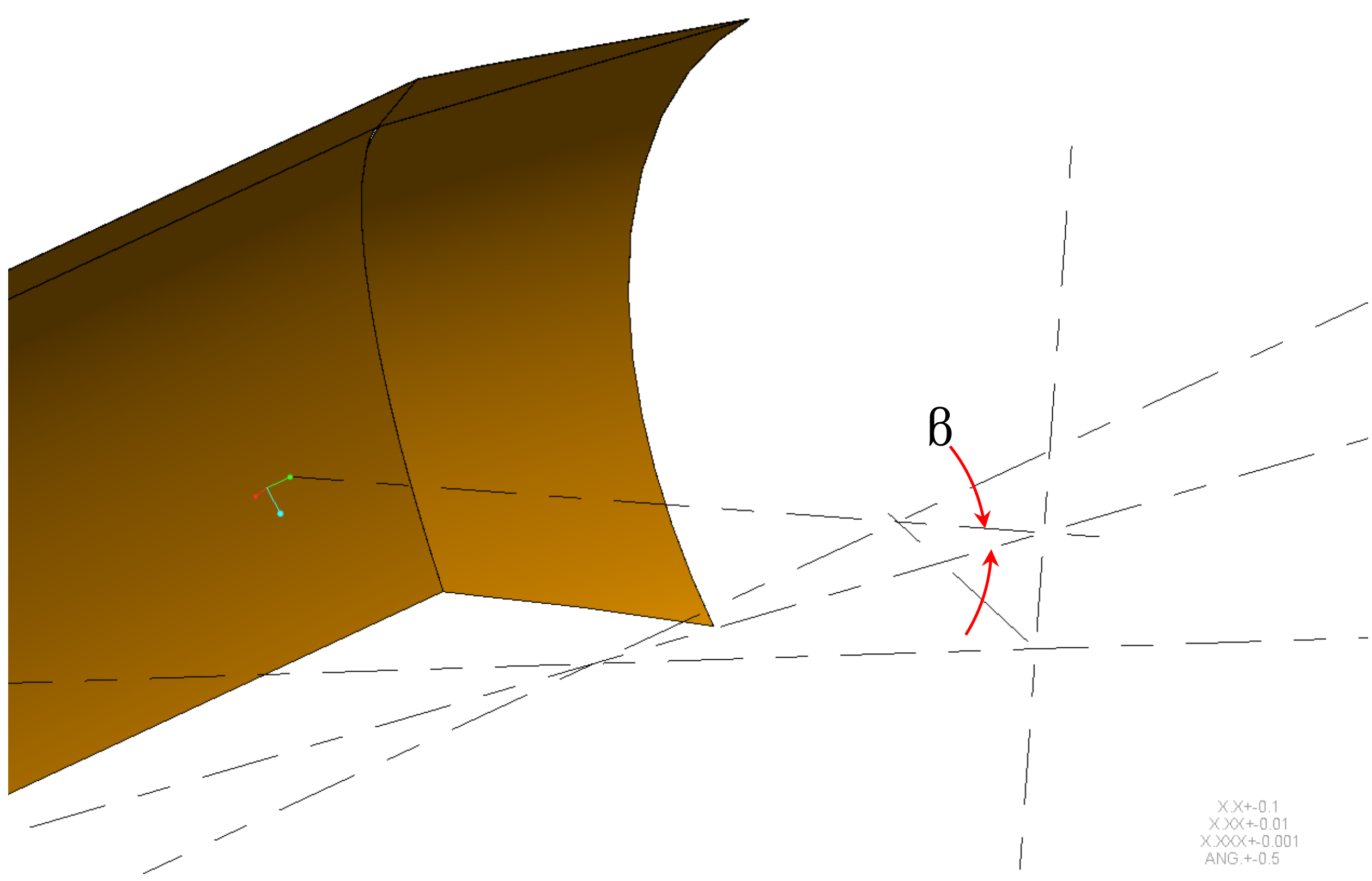

Fig. 23: β = Angle of intersection between the primary cylinder axis and an axis lying in the vertical orthogonal plane

Now refer to Fig. 24 for a definition of the second parameter required of the interior curve equations. The angle α is defined as the angle of rotation required for the interior curves created by the intersection of the primary and secondary cylindrical sections. This angle is measured on the tetrahedron as the sum of σ and the angle measured between the wing horizontal plane and the plane normal to the flat surface at the top of the primary cylinder surface as shown. This interior angle is in effect σ + (90°–γ) where σ is the angle between the primary cylinder and wing horizontal planes and (90°-γ) is the angle between the tertiary orthogonal plane and the horizontal orthogonal plane. It will be designated as φ - 90° in order to assign the correct sign to the angular value required to orient the curves in the proper direction in accordance with the choices made for θ and ω (wing angle and rotation).

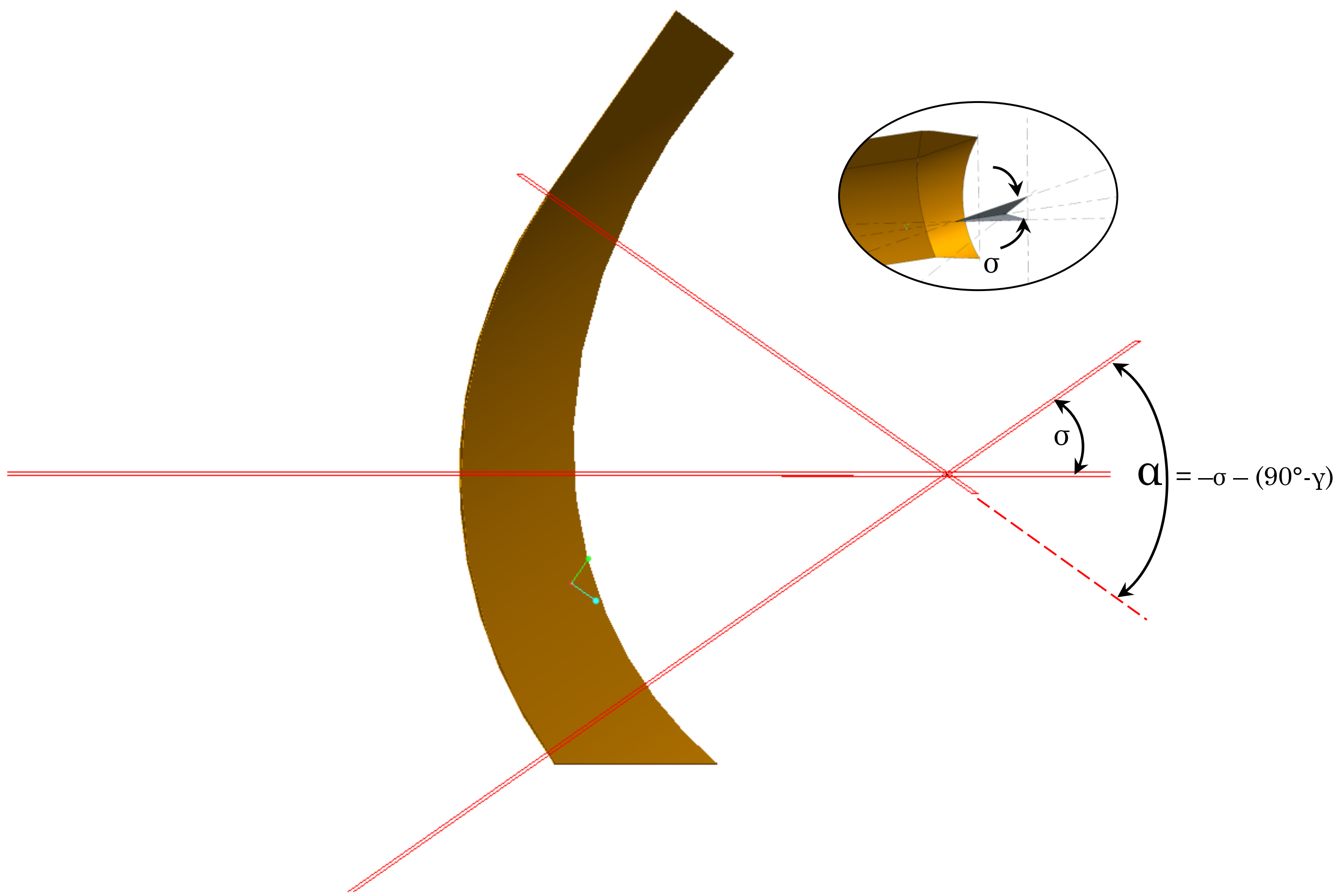

Fig. 24: α = [–σ – (90°–γ)] = Rotation of the cut between the primary cylinder and wing cylinder surfaces

We will now analyze the parameters required of the curve equation that creates the wing outer end cut. This curve is unique in that it is a sinusoidal curve in the flat state which constrains the outer edge of the wing section to lie precisely on the vertical orthogonal plane when the geometry is formed into the 3-D state. The first of the two angles required to write the equation for the wing end curve is designated as ρ, and as it turns out this angle is actually equal to δ, or the angle of tilt between the primary cylinder axis and the wing cylinder axis which was calculated for the chosen values of θ and ω (wing angle and wing rotation). Please refer to Fig. 25 for a visual depiction of this parameter.

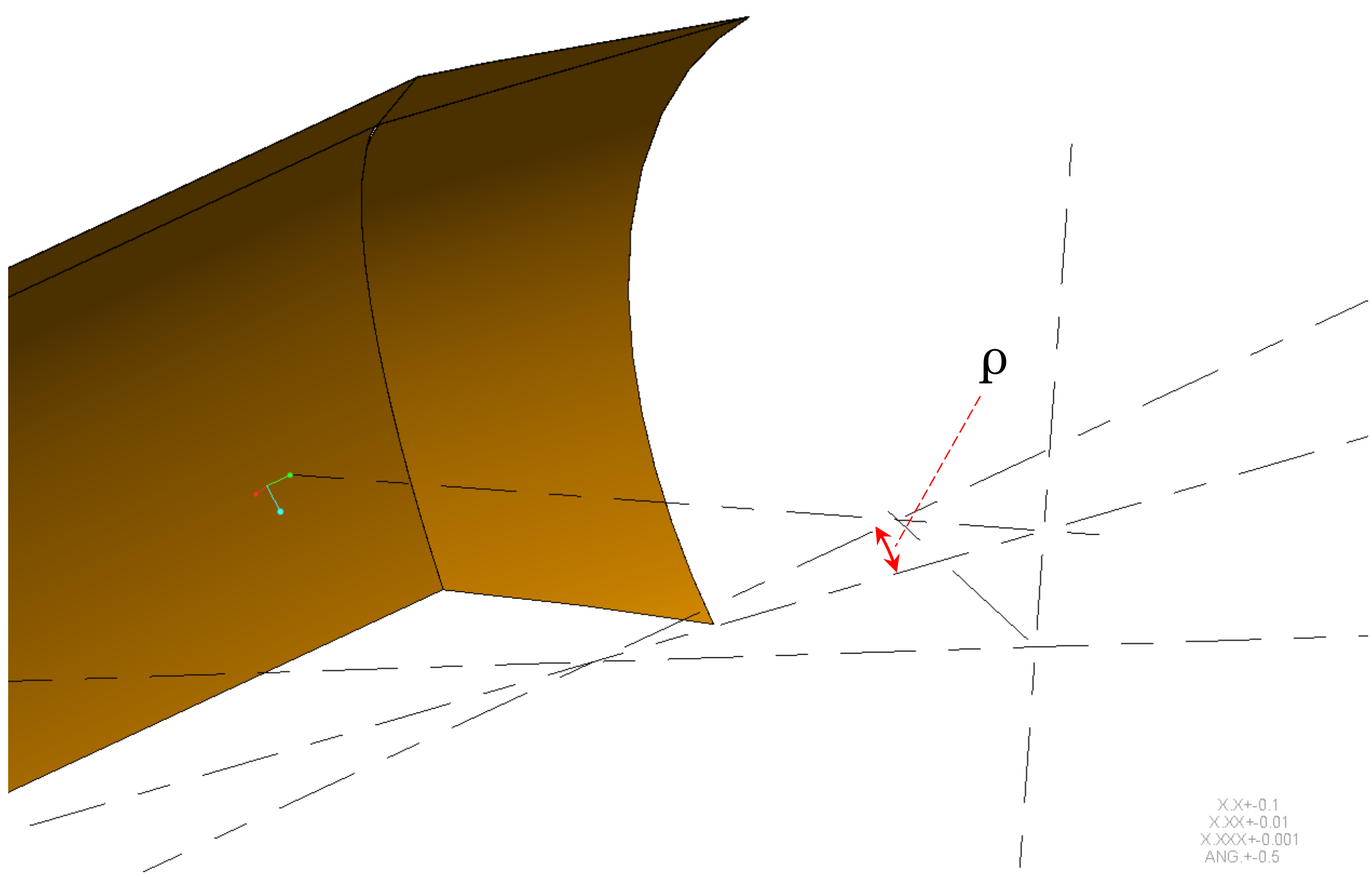

Fig. 25: ρ = Angle of cut for the wing end curve

The second of the two angles required of the sinusoidal curve equation to perform the wing outer end cut is the angle of rotation of the wing end curve with respect to the primary cylinder axis. Please see Fig. 26 for a visual depiction of this angle. This angle is designated as φ, and it is measured as the interior angle in between the wing horizontal plane and the plane of the flat upper portion of the wing, or:

$$\gamma - \sigma$$

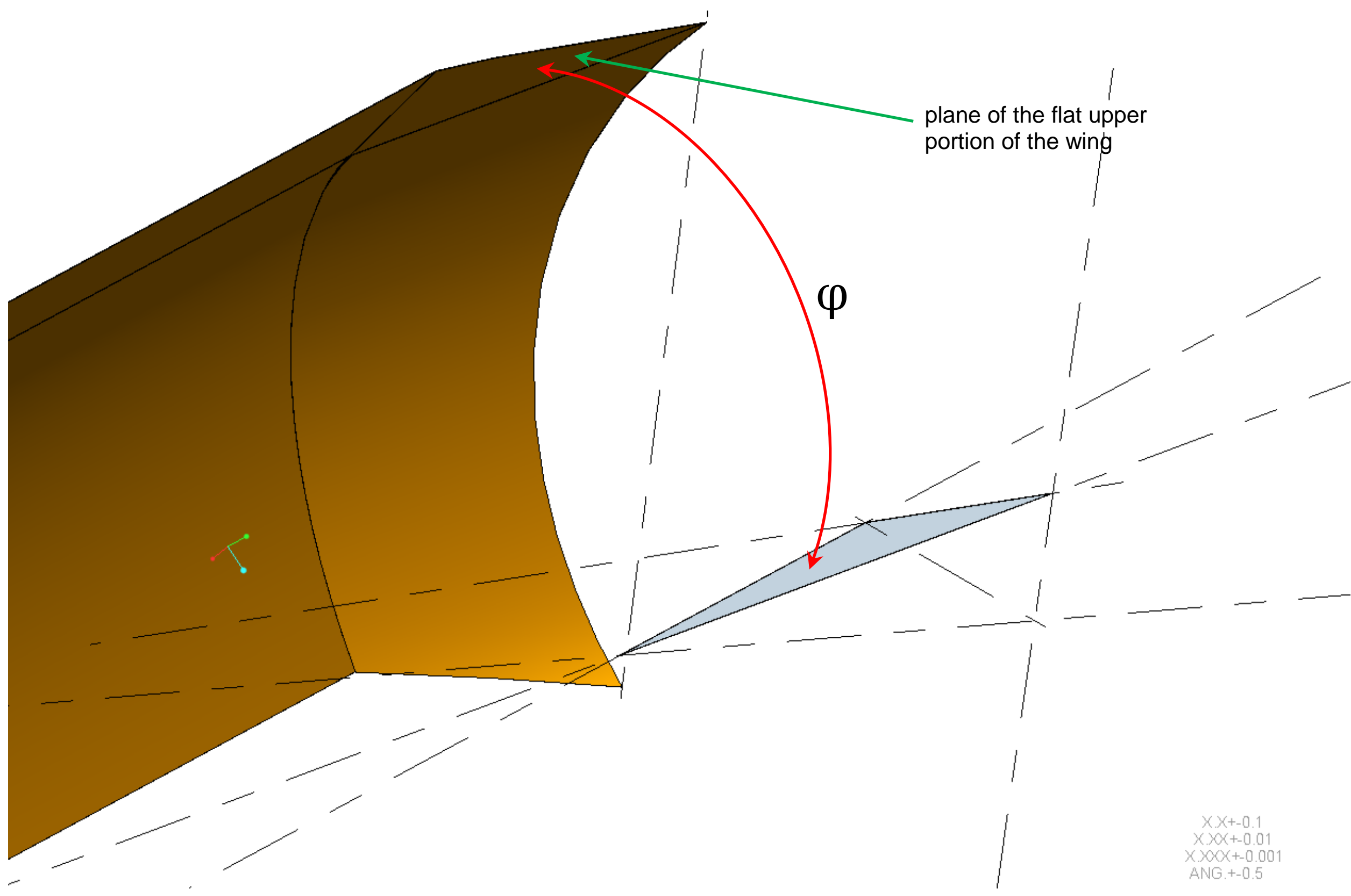

Fig. 26: $\varphi$ = Angle of rotation for the wing end curve

We'll now analyze the curve required to create the wing bottom cut. This curve is also unique in that it is a sinusoidal curve in the flat state which constrains the bottom edge of the wing section to lie precisely on the horizontal orthogonal plane when the geometry is formed into the 3-D state. The first angle required to write the equation for the curve which performs the wing bottom cut is designated as $\upsilon$, and it is measured on the tetrahedron as the angle between the wing cylindrical axis and an axis perpendicular to the horizontal orthogonal plane. Please see Fig. 27 for a visual depiction of this angular value.

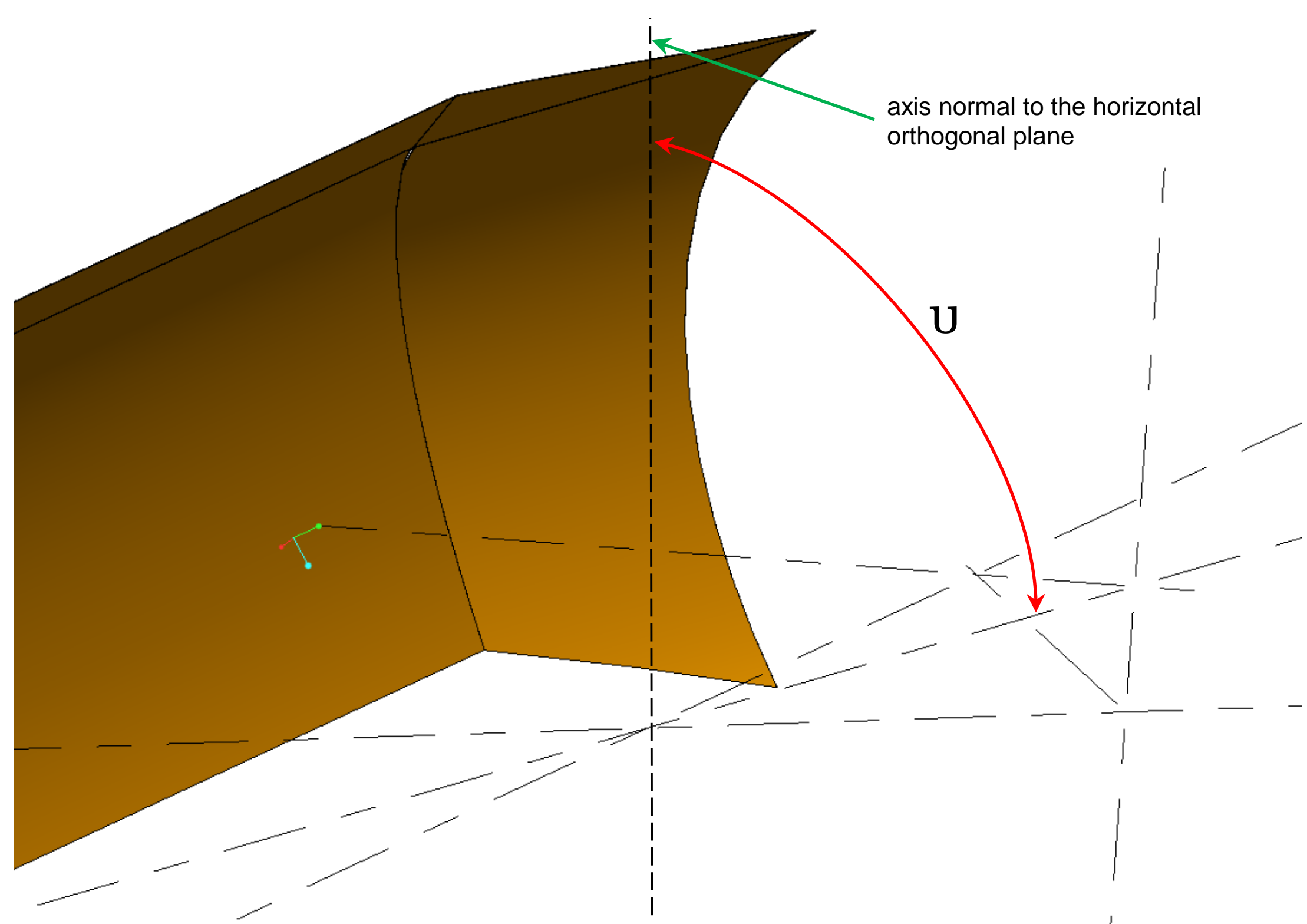

Fig. 27: υ = Angle of cut for the wing bottom cut

The second angle that is required of the sinusoidal curve equation to execute the wing bottom cut is the angle of rotation of the wing bottom curve with respect to the primary cylinder axis. Please see Fig. 28 for a visual depiction of this angle. This angle is designated as ψ, and it is measured as the interior angle in between the vertical plane formed in the tetrahedron (which is perpendicular to the horizontal orthogonal plane) and the plane of the flat upper portion of the wing.

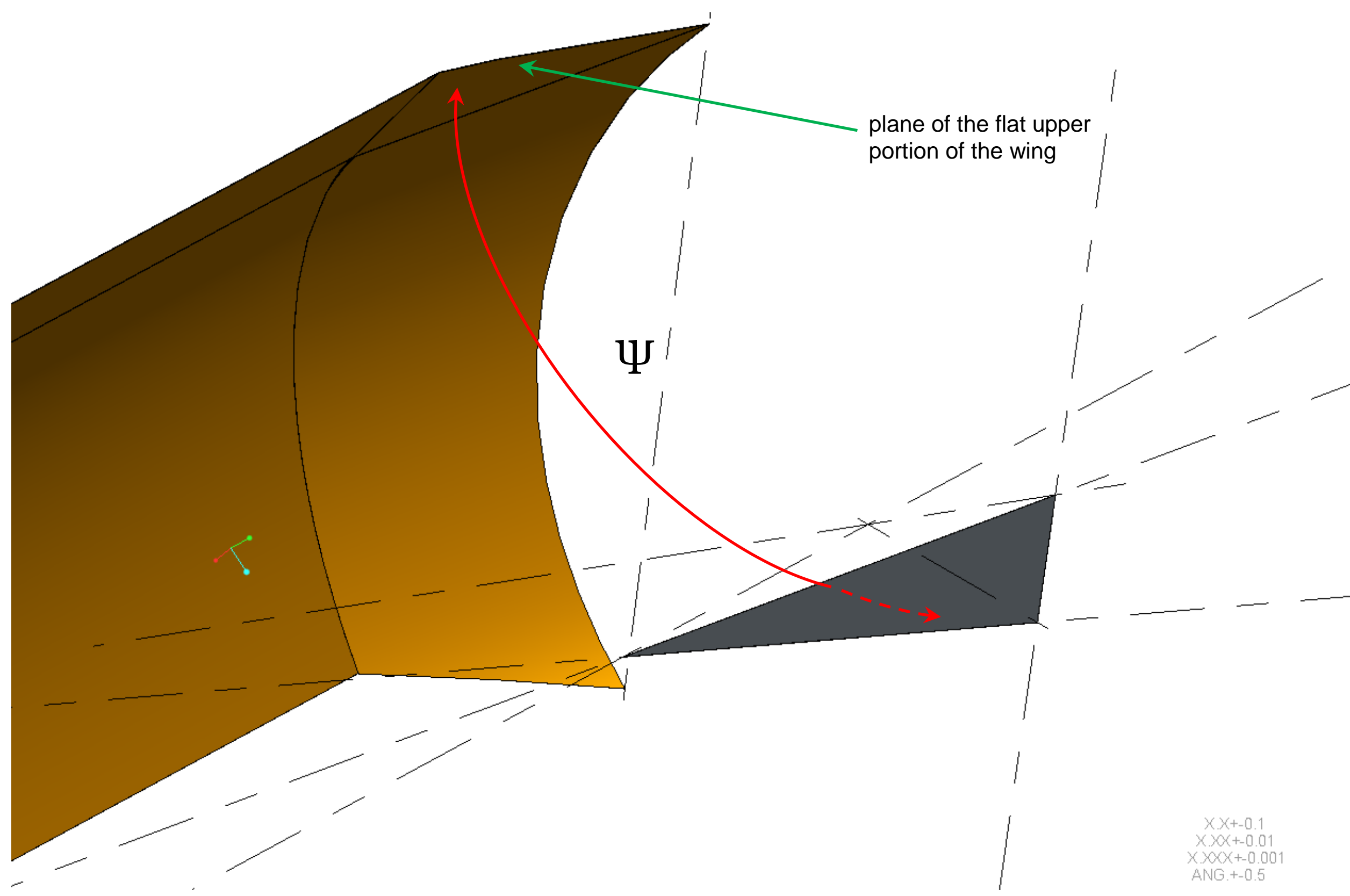

Fig. 28: ψ = Angle of rotation for the wing bottom cut

This concludes the definition of the parameters that are required as inputs for the sinusoidal curve cutting equations to create the geometry in its flat state. As was intimated earlier, a further calculation is required to determine the slope of the line across which the interior cutting curve of the primary cylinder surface section is mirrored in the flat state geometry to produce the wing interior curve. Please refer to Fig. 29 for a visual definition of this line, the slope of which is designated as μ. This slope is calculated by taking the derivative of Eq. 1 written for the interior curves, setting $x = 0$, and then entering the parameters α and β calculated above. The resultant curve f(u,v) for the wing interior cut is effectively mirrored across this line of reflection by multiplying Eq. 1 for the primary cylinder interior cut, or f(x,y), by the matrix operator as follows:

$$f(u,v) = f(x,y) * \begin{Bmatrix} cos\mu & sin\mu \\ -sin\mu & cos\mu \end{Bmatrix}$$

where μ is the slope of the mirror line.

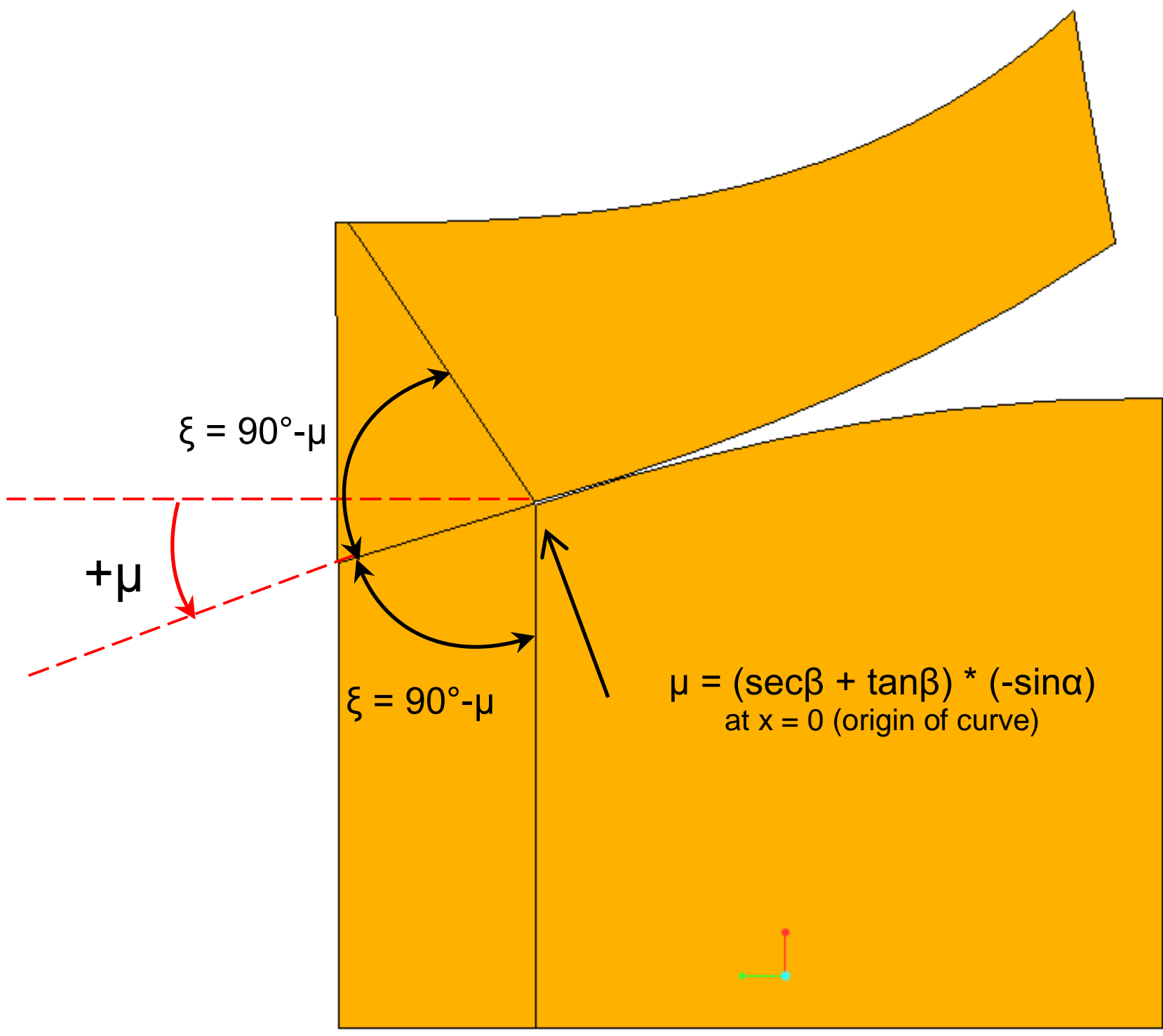

Fig. 29: Calculation of the slope of the mirror line for reflecting the interior cut curve on the primary cylinder to produce the interior cut on the wing

One final calculation is required to trim the top edge of the flat upper section of the wing so that that this edge will lie entirely within the tertiary orthogonal plane that passes through the top edge of the flat upper section of the primary cylinder after the rolling and bending operations are completed to form the geometry into its 3-D state. Please see Fig. 30 for a review the definition of the tertiary orthogonal plane. The calculations used to determine the angle for the wing top cut are illustrated in Fig. 31. Fig. 32 illustrates the general profile of the calculated trim angle.

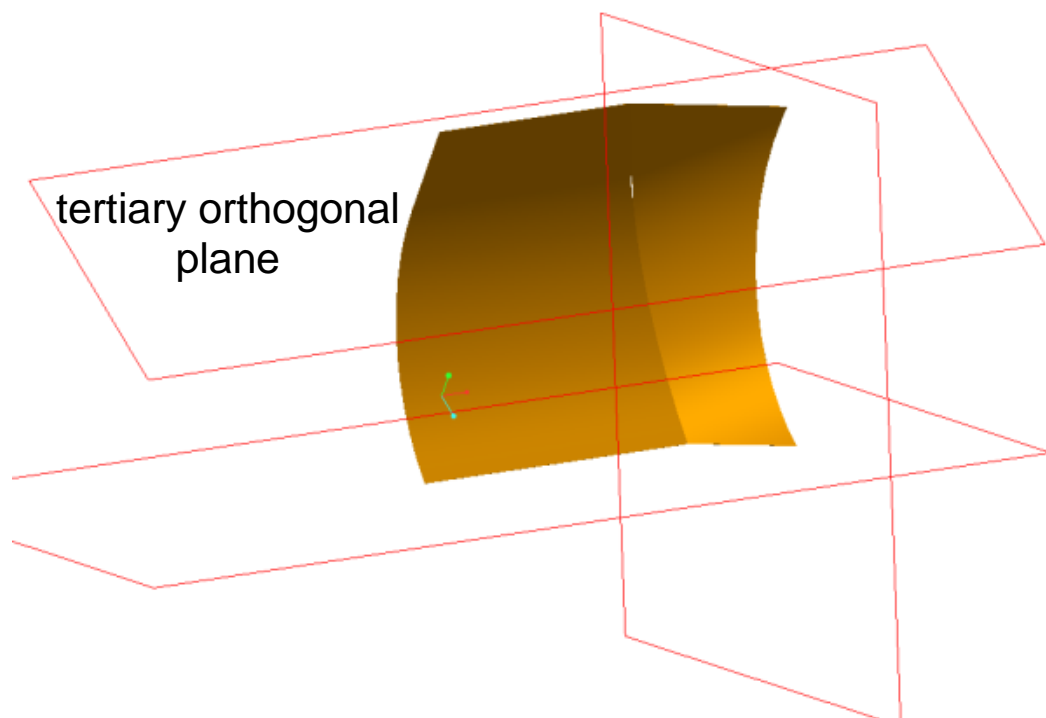

Fig. 30: Illustration of the tertiary orthogonal plane

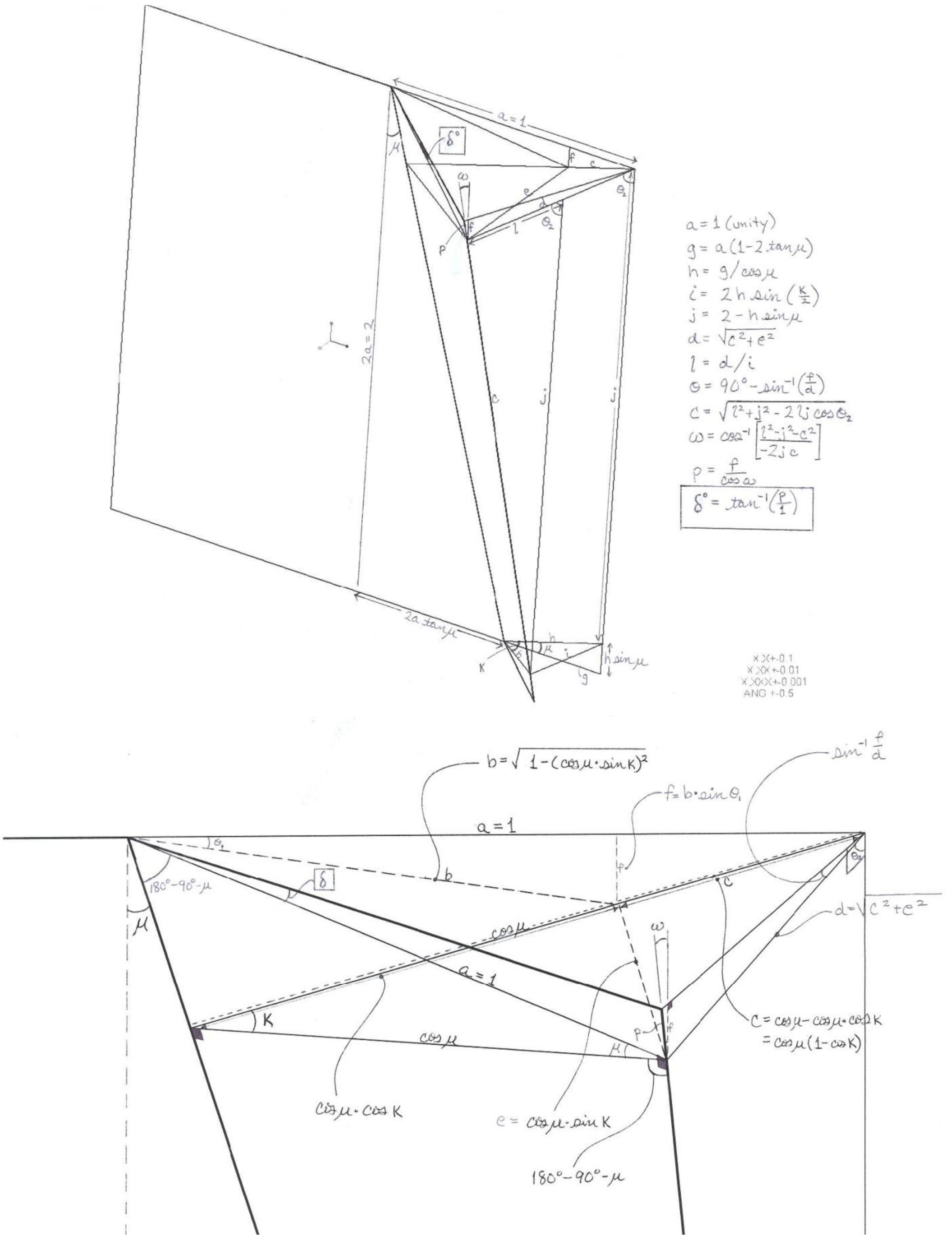

Fig. 31: Calculation of the trim angle δ required for the top edge of the wing flat section to enable alignment to the tertiary orthogonal plane in the 3-D state

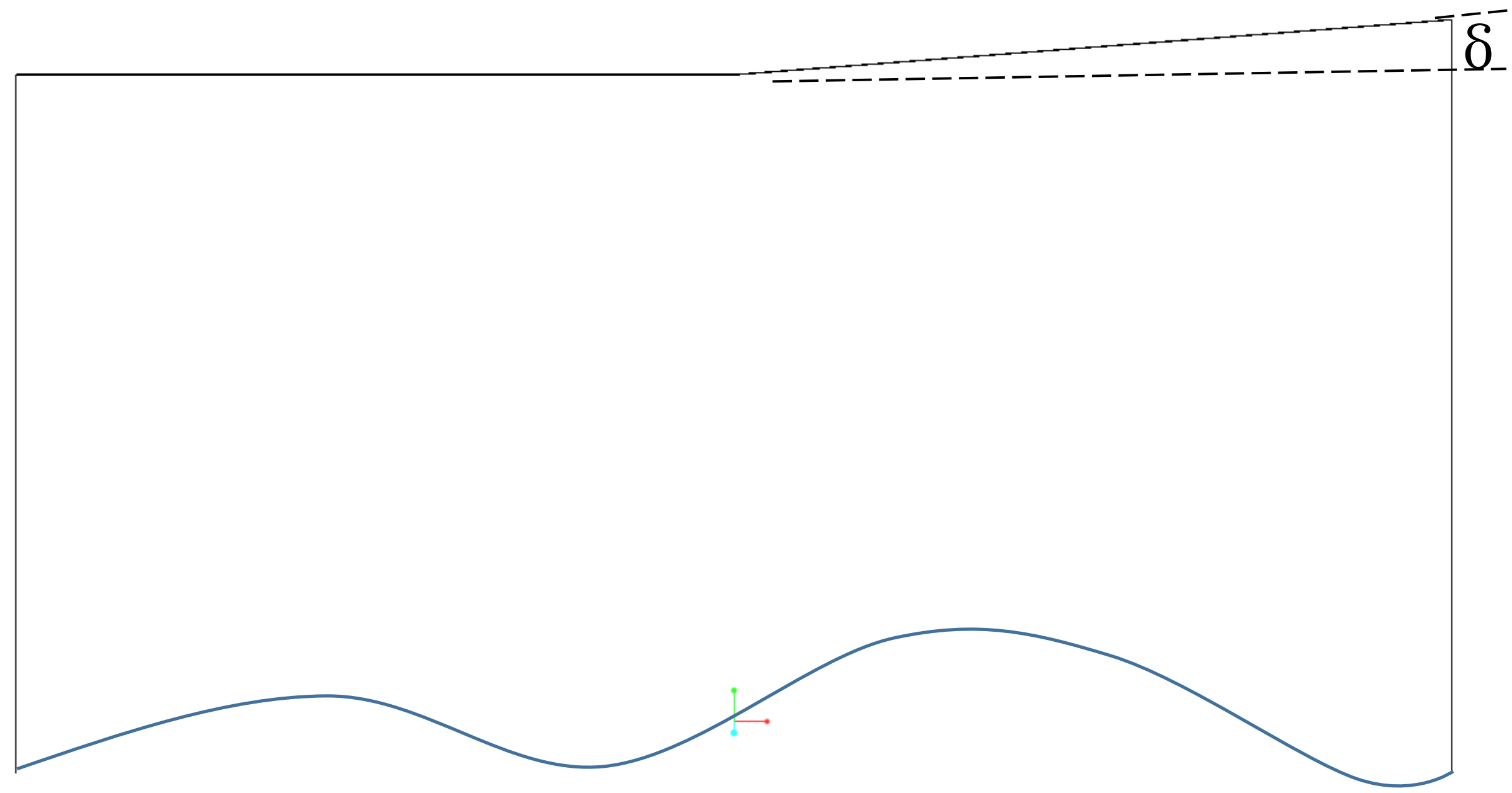

Fig. 32: General profile of the trim angle for the top edge of the wing flat upper surface

After all wing interior and exterior cuts are made using the sinusoidal curve equations and the trim angle calculation, then the next operation is to roll the primary cylinder surface and wing surface sections to the radius which was initially chosen and used in all the curve equations up to this point. It is important to note that the primary cylinder and wing surface sections should be rolled about their own cylindrical axes and that the curved sections terminate at the tangent lines that divide the flat upper portions from the lower curved sections of the primary and wing cylinder surfaces. Please see the left-hand view of Fig. 33 for a depiction of the rolling process.

Finally, a bend angle is required to close the gap remaining between the interior cutting curves after the cutting and rolling operations are complete. Calculating this angle is accomplished by constructing a diagram of the angular intersection of the primary cylinder and wing cylinder axes and then using the law of cosines in two successive applications to arrive at this angle. Please see the right-hand view in Fig. 33 for a visual depiction of the bend angle and Fig. 34 for the bend angle calculation details.

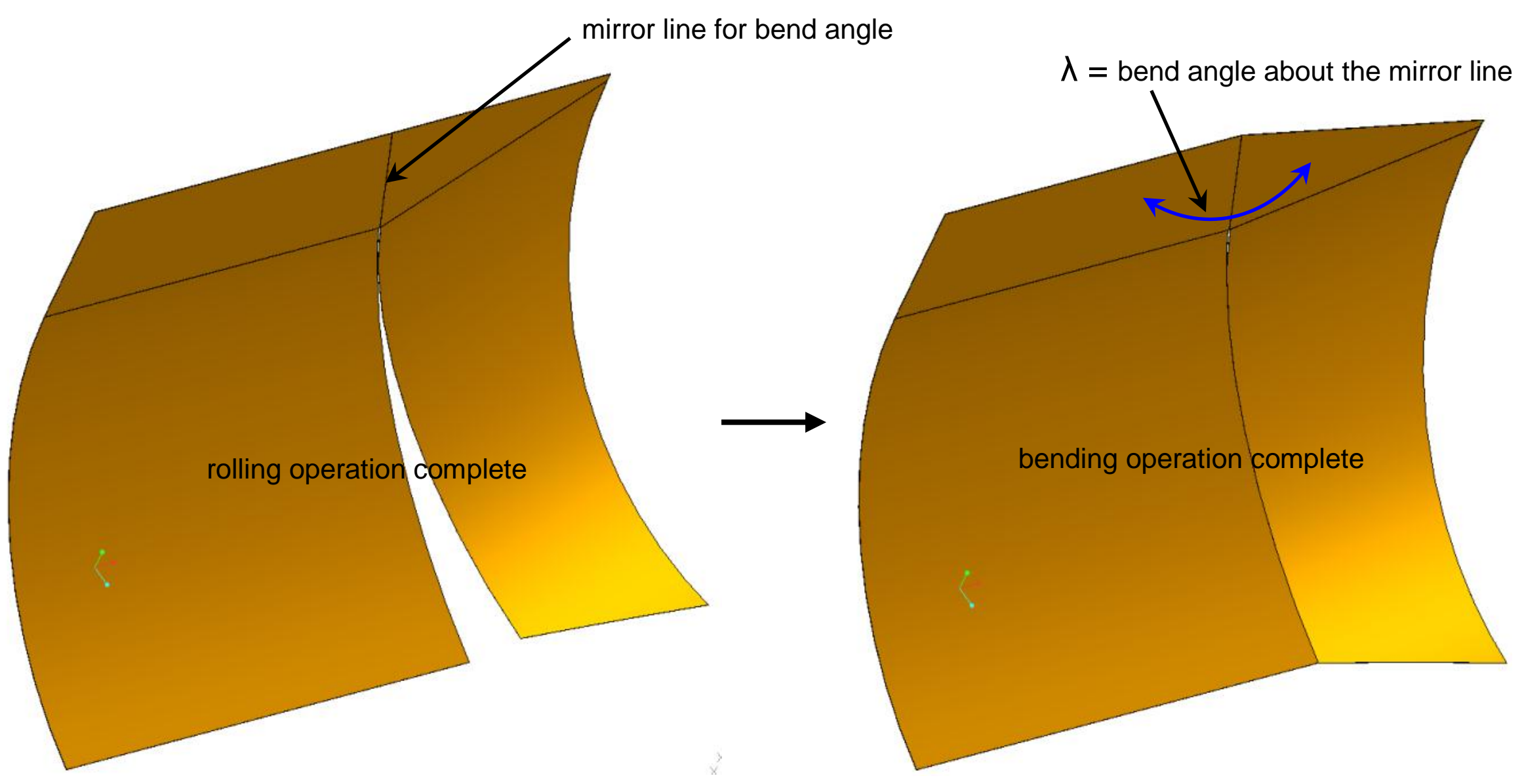

Fig. 33: Bend angle between the flat upper sections of the primary cylinder and wing sections

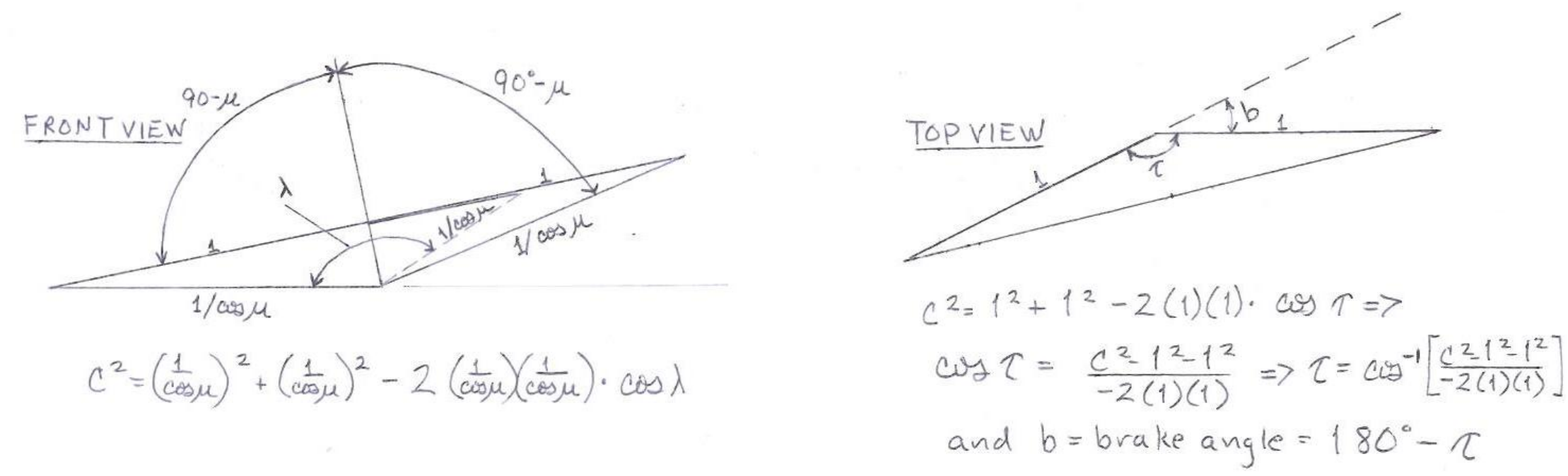

Fig. 34: Calculation of the bend angle: λ = 180° - τ

This completes the definition of a closed-form solution to create an orthogonal cross section of multivariate cylinder surface intersection geometry from the flat state. In conclusion, the method outlined in this research paper for creating cylinder surface intersection geometry from the flat state could possibly be extended to airfoil wings, intersecting pipe sections, cartography, or any application which would require a mathematical approach for solving such geometry. The method outlined in this research paper could also be adapted to create continuous cylinder surface intersections from a single flat blank as illustrated in Fig. 34.

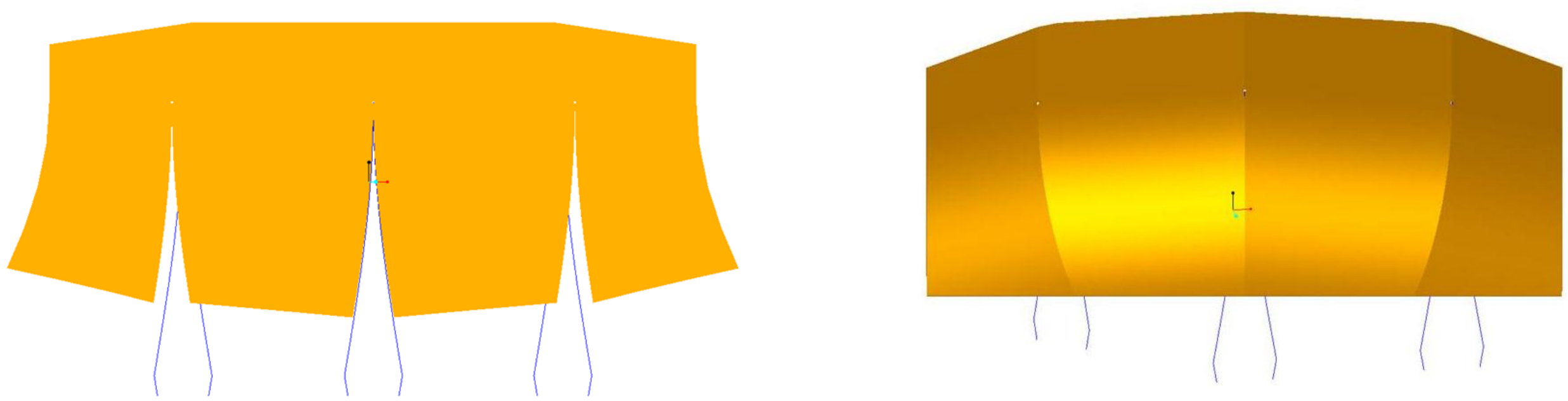

Fig. 34: Multiple cylinder surface intersections created from a one-piece blank

The author anticipates that a dimensionless number (similar to a Reynolds number) could be calculated from parameters extracted from the tetrahedron which would aid in predicting fluid flow patterns across the intersecting cylinder surface sections in different flow situations. Further research, of course, would be required to ascertain this claim for the variable geometry.

To conclude, please see Fig. 35 for an illustration of a computer spreadsheet created by the author which automates the calculation of the many parameters and trigonometric calculations required to create the multivariate cylinder surface geometry from the flat state as the primary and secondary angles of wing rotation are iterated to vary the geometry within the bounds of the orthogonal planes. The work required to produce the parameters required of the sinusoidal equations, as well as the other trigonometric calculations needed to create the geometry, is reduced significantly through the use of a computer application. For example, in the spreadsheet illustrated in Fig. 35, one has only to enter arbitrarily-chosen values of the primary and secondary directions of wing rotation in order to produce all the calculations necessary for iterating the geometry.

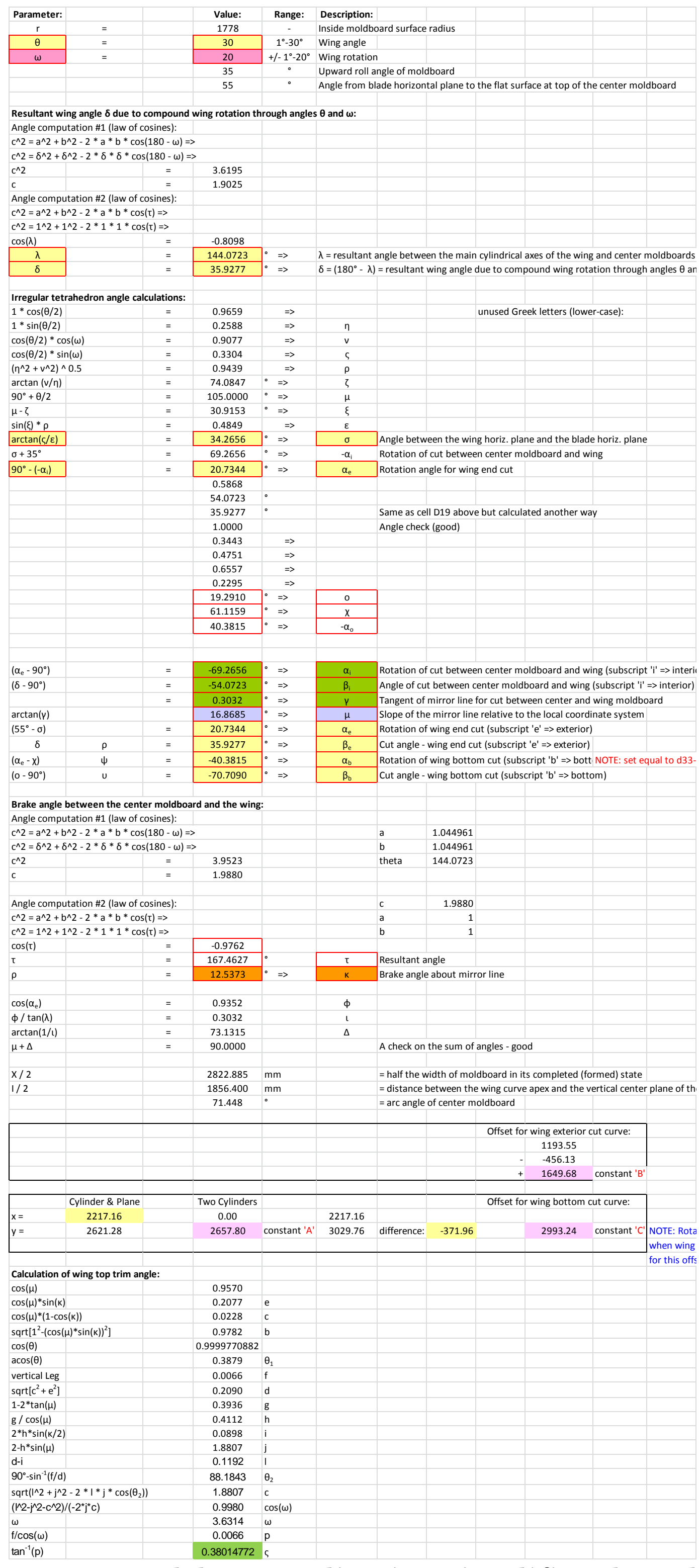

| Parameter: | | | Value: | Range: | Description: | | | | | | |
|---|---|---|---|---|---|---|---|---|---|---|---|
| r | = | | 1778 | - | Inside moldboard surface radius | | | | | | |
| θ | = | | 30 | 1°-30° | Wing angle | | | | | | |
| ω | = | | 20 | +/- 1°-20° | Wing rotation | | | | | | |
| | | | 35 | ° | Upward roll angle of moldboard | | | | | | |
| | | | 55 | ° | Angle from blade horizontal plane to the flat surface at top of the center moldboard | | | | | | |
| | | | | | | | | | | | |
| **Resultant wing angle δ due to compound wing rotation through angles θ and ω:** | | | | | | | | | | | |
| Angle computation #1 (law of cosines): | | | | | | | | | | | |
| c^2 = a^2 + b^2 - 2 * a * b * cos(180 - ω) => | | | | | | | | | | | |
| c^2 = δ^2 + δ^2 - 2 * δ * δ * cos(180 - ω) => | | | | | | | | | | | |
| c^2 | | = | 3.6195 | | | | | | | | |
| c | | = | 1.9025 | | | | | | | | |
| Angle computation #2 (law of cosines): | | | | | | | | | | | |
| c^2 = a^2 + b^2 - 2 * a * b * cos(τ) => | | | | | | | | | | | |
| c^2 = 1^2 + 1^2 - 2 * 1 * 1 * cos(τ) => | | | | | | | | | | | |
| cos(λ) | | = | -0.8098 | | | | | | | | |
| λ | | = | 144.0723 | ° => | λ = resultant angle between the main cylindrical axes of the wing and center moldboards | | | | | | |
| δ | | = | 35.9277 | ° => | δ = (180° - λ) = resultant wing angle due to compound wing rotation through angles θ an | | | | | | |
| | | | | | | | | | | | |
| **Irregular tetrahedron angle calculations:** | | | | | | | | | | | |
| 1 * cos(θ/2) | | = | 0.9659 | => | | | | unused Greek letters (lower-case): | | | |
| 1 * sin(θ/2) | | = | 0.2588 | => | η | | | | | | |
| cos(θ/2) * cos(ω) | | = | 0.9077 | => | ν | | | | | | |
| cos(θ/2) * sin(ω) | | = | 0.3304 | => | ς | | | | | | |
| (η^2 + ν^2) ^ 0.5 | | = | 0.9439 | => | ρ | | | | | | |
| arctan (ν/η) | | = | 74.0847 | ° => | ζ | | | | | | |
| 90° + θ/2 | | = | 105.0000 | ° => | μ | | | | | | |
| μ - ζ | | = | 30.9153 | ° => | ξ | | | | | | |
| sin(ξ) * ρ | | = | 0.4849 | => | ε | | | | | | |
| arctan(ς/ε) | | = | 34.2656 | ° => | σ | Angle between the wing horiz. plane and the blade horiz. plane | | | | | |
| σ + 35° | | = | 69.2656 | ° => | $-\alpha_i$ | Rotation of cut between center moldboard and wing | | | | | |
| 90° - ($-\alpha_i$) | | = | 20.7344 | ° => | $\alpha_e$ | Rotation angle for wing end cut | | | | | |
| | | | 0.5868 | | | | | | | | |
| | | | 54.0723 | ° | | | | | | | |
| | | | 35.9277 | ° | | Same as cell D19 above but calculated another way | | | | | |
| | | | 1.0000 | | | Angle check (good) | | | | | |
| | | | 0.3443 | => | | | | | | | |
| | | | 0.4751 | => | | | | | | | |
| | | | 0.6557 | => | | | | | | | |
| | | | 0.2295 | => | | | | | | | |
| | | | 19.2910 | ° => | ο | | | | | | |
| | | | 61.1159 | ° => | χ | | | | | | |
| | | | 40.3815 | ° => | $-\alpha_o$ | | | | | | |
| | | | | | | | | | | | |
| | | | | | | | | | | | |
| ($\alpha_e$ - 90°) | | = | -69.2656 | ° => | $\alpha_i$ | Rotation of cut between center moldboard and wing (subscript 'i' => interi | | | | | |
| (δ - 90°) | | = | -54.0723 | ° => | $\beta_i$ | Angle of cut between center moldboard and wing (subscript 'i' => interior) | | | | | |
| | | = | 0.3032 | ° => | γ | Tangent of mirror line for cut between center and wing moldboard | | | | | |
| arctan(γ) | | | 16.8685 | ° => | μ | Slope of the mirror line relative to the local coordinate system | | | | | |
| (55° - σ) | | = | 20.7344 | ° => | $\alpha_e$ | Rotation of wing end cut (subscript 'e' => exterior) | | | | | |
| δ | ρ | = | 35.9277 | ° => | $\beta_e$ | Cut angle - wing end cut (subscript 'e' => exterior) | | | | | |
| ($\alpha_e$ - χ) | ψ | = | -40.3815 | ° => | $\alpha_b$ | Rotation of wing bottom cut (subscript 'b' => bott | | | | NOTE: set equal to d33- | |
| (ο - 90°) | υ | = | -70.7090 | ° => | $\beta_b$ | Cut angle - wing bottom cut (subscript 'b' => bottom) | | | | | |
| | | | | | | | | | | | |
| **Brake angle between the center moldboard and the wing:** | | | | | | | | | | | |
| Angle computation #1 (law of cosines): | | | | | | | | | | | |
| c^2 = a^2 + b^2 - 2 * a * b * cos(180 - ω) => | | | | | | a | 1.044961 | | | | |
| c^2 = δ^2 + δ^2 - 2 * δ * δ * cos(180 - ω) => | | | | | | b | 1.044961 | | | | |
| c^2 | | = | 3.9523 | | | theta | 144.0723 | | | | |
| c | | = | 1.9880 | | | | | | | | |
| | | | | | | | | | | | |
| Angle computation #2 (law of cosines): | | | | | | c | 1.9880 | | | | |
| c^2 = a^2 + b^2 - 2 * a * b * cos(τ) => | | | | | | a | 1 | | | | |
| c^2 = 1^2 + 1^2 - 2 * 1 * 1 * cos(τ) => | | | | | | b | 1 | | | | |
| cos(τ) | | = | -0.9762 | | | | | | | | |
| τ | | = | 167.4627 | ° | τ | Resultant angle | | | | | |
| ρ | | = | 12.5373 | ° => | κ | Brake angle about mirror line | | | | | |
| | | | | | | | | | | | |
| cos($\alpha_e$) | | = | 0.9352 | | ϕ | | | | | | |
| ϕ / tan(λ) | | = | 0.3032 | | ι | | | | | | |
| arctan(1/ι) | | = | 73.1315 | | Δ | | | | | | |
| μ + Δ | | = | 90.0000 | | | A check on the sum of angles - good | | | | | |
| | | | | | | | | | | | |
| X / 2 | | | 2822.885 | mm | | = half the width of moldboard in its completed (formed) state | | | | | |
| I / 2 | | | 1856.400 | mm | | = distance between the wing curve apex and the vertical center plane of th | | | | | |
| | | | 71.448 | ° | | = arc angle of center moldboard | | | | | |
| | | | | | | | | | | | |
| | | | | | | | | | Offset for wing exterior cut curve: | | |
| | | | | | | | | | 1193.55 | | |
| | | | | | | | | - | -456.13 | | |
| | | | | | | | | + | 1649.68 | constant 'B' | |
| | | | | | | | | | | | |
| | Cylinder & Plane | | Two Cylinders | | | | | | Offset for wing bottom cut curve: | | |
| x = | 2217.16 | | 0.00 | | 2217.16 | | | | | | |
| y = | 2621.28 | | 2657.80 | constant 'A' | 3029.76 | difference: | -371.96 | | 2993.24 | constant 'C' | NOTE: Rota |
| | | | | | | | | | | | when wing |
| | | | | | | | | | | | for this offs |
| **Calculation of wing top trim angle:** | | | | | | | | | | | |
| cos(μ) | | | 0.9570 | | | | | | | | |
| cos(μ)*sin(κ) | | | 0.2077 | e | | | | | | | |
| cos(μ)*(1-cos(κ)) | | | 0.0228 | c | | | | | | | |
| sqrt[$1^2$-(cos(μ)*sin(κ))$^2$] | | | 0.9782 | b | | | | | | | |
| cos(θ) | | | 0.9999770882 | | | | | | | | |
| acos(θ) | | | 0.3879 | $\theta_1$ | | | | | | | |
| vertical Leg | | | 0.0066 | f | | | | | | | |
| sqrt[$c^2$ + $e^2$] | | | 0.2090 | d | | | | | | | |
| 1-2*tan(μ) | | | 0.3936 | g | | | | | | | |
| g / cos(μ) | | | 0.4112 | h | | | | | | | |
| 2*h*sin(κ/2) | | | 0.0898 | i | | | | | | | |
| 2-h*sin(μ) | | | 1.8807 | j | | | | | | | |
| d-i | | | 0.1192 | l | | | | | | | |
| 90°-$\sin^{-1}$(f/d) | | | 88.1843 | $\theta_2$ | | | | | | | |
| sqrt(l^2 + j^2 - 2 * l * j * cos($\theta_2$)) | | | 1.8807 | c | | | | | | | |
| (l^2-j^2-c^2)/(-2*j*c) | | | 0.9980 | cos(ω) | | | | | | | |
| ω | | | 3.6314 | ω | | | | | | | |
| f/cos(ω) | | | 0.0066 | p | | | | | | | |
| $\tan^{-1}$(p) | | | 0.38014772 | ς | | | | | | | |

Fig. 35: Computer spreadsheet application simplifies the calculation of the parameters necessary for creating the multivariate cylinder intersection geometry as the primary and secondary angles of rotation are iterated to vary the geometry within the bounds of the orthogonal planes